\documentclass[preprints,article,accept,moreauthors]{Definitions/mdpi}

\firstpage{1}
\pubvolume{1}
\issuenum{1}
\articlenumber{0}
\pubyear{2026}
\copyrightyear{2026}
\datereceived{}
\daterevised{}
\dateaccepted{}
\datepublished{}

\Title{Effects of Soret Diffusion and Radiative Heat Loss on the Evolution of Buoyant Flame Kernels in Ultra-Lean Hydrogen-Air Mixture}

\Author{Ivan S. Yakovenko $^{1,}$*\orcidA{}, Alexey D. Kiverin $^{1}$\orcidB{}}
\AuthorNames{Ivan S. Yakovenko, Alexey D. Kiverin}

\address[1]{%
	$^{1}$ \quad Joint Institute for High Temperatures of the Russian Academy of Sciences, Izhorskaya St. 13 Bldg. 2, Moscow 125412, Russia;}

\corres{Correspondence: yakovenko.ivan@bk.ru}

\abstract{Ultra-lean hydrogen flames under terrestrial gravity are governed by a coupled interaction among preferential diffusion, thermal diffusion, heat loss, and self-induced convection. This study numerically examines combustion in a quiescent 6~vol.\% H$_2$--air mixture using detailed chemistry and a low-Mach-number formulation. A complete calculations set was considered, with Soret diffusion and optically thin radiative heat loss independently enabled and disabled. One-dimensional spherical calculations were used to isolate the initial post-ignition flame kernel growth, while two-dimensional planar and axisymmetric simulations described its subsequent buoyant rise, deformation, and breakup. Over the analyzed interval, the spherical flame-front radius followed $R_f^2\approx Kt$ rather than constant-speed expansion. Soret diffusion increased the effective growth coefficient $K$, whereas radiation reduced it. The axisymmetric calculations reproduced the experimentally measured leading-point trajectory substantially better than the planar formulation. Soret diffusion produced larger, faster-rising kernels and maintained a more nearly circular upper cap, whereas radiation had a weaker effect on trajectory but increased relative lateral flattening. In all cases, a toroidal vortex stretched the flame segment and caused local extinction and fragmentation. Soret diffusion delayed breakup, while radiation advanced it; their combined effect on breakup time was nearly compensating. The results show that Soret transport and radiation primarily alter kernel growth and resistance to vortex-induced extinction, while the qualitative breakup pathway remains hydrodynamically controlled.}

\begin{document}
	
	\section{Introduction}
	
	Hydrogen is increasingly viewed as an important energy carrier for decarbonizing transport, power generation, energy storage, and propulsion systems. Although hydrogen combustion produces no carbon dioxide at the point of use, the wider adoption of hydrogen technologies requires reliable methods for preventing and mitigating accidental releases, fires, and explosions. Leaks from storage vessels, pipelines, valves, refuelling equipment, and other facilities may gradually form flammable hydrogen--air mixtures in confined or semi-confined spaces \cite{makhviladze1998csat, makhviladze2002potci}. The associated hazard is not limited to highly reactive mixtures close to stoichiometric conditions. Imperfect mixing, ventilation, and natural stratification can produce large lean regions alongside locally hydrogen-rich layers. A slowly propagating flame initiated in an ultra-lean region may therefore act as a moving ignition source, transferring heat and reactive species toward more combustible areas of the gas cloud. Recent studies of hydrogen dispersion, pipeline leakage, jet fires, explosions, and combustion modelling highlight the importance of multiscale flame behaviour in modern fire-risk assessment \cite{Yakovenko2023Fire, Lin2024Fire}.
	
	A related hazard arises from thermal runaway in lithium-ion batteries. Battery failure can release a multicomponent flammable vent gas containing hydrogen, carbon monoxide, methane, ethylene, and other combustible species, with the relative proportions depending on battery chemistry and operating conditions. When these gases accumulate in battery enclosures, electric vehicles, energy-storage installations, or poorly ventilated rooms, they may form strongly non-uniform combustible clouds. The dynamics and stability limits of weakly reactive hydrogen-containing flames are therefore relevant not only to dedicated hydrogen infrastructure, but also to energy storage facilities. Recent work has examined electrolyte oxidation and fire-gas production during lithium-ion battery thermal runaway \cite{Tian2025Fire}, while a dedicated review identifies CO, CO$_2$, H$_2$, CH$_4$, C$_2$H$_4$, and other hydrocarbons among the main components of battery vent gas \cite{Qiu2023Batteries}.
	
	The importance of buoyancy-dominated combustion is not unique to hydrogen. Ammonia is also being considered as a carbon-free fuel and hydrogen carrier, but its low reactivity and small laminar burning velocity make ammonia--air flames particularly sensitive to gravitational effects. Experiments have shown that buoyancy can strongly modify ammonia-flame shape and apparent propagation characteristics, especially at elevated pressure or near the flammability limits, where slowly expanding flames may develop a pronounced cap-like structure \cite{Alvarez2024Ammonia, babkin1982cesw}. These observations suggest that the interaction between weak chemical reactivity, natural convection, flame stretch, and heat loss is a broader problem affecting several prospective carbon-free fuels.
	
	Combustion in ultra-lean hydrogen--air mixtures differs fundamentally from conventional premixed deflagration. Near the lower flammability limit, the low rate of chemical energy release and the high diffusivity of molecular hydrogen favour flame structures sustained by diffusion of the deficient reactant toward the reaction zone. Under microgravity conditions, this behaviour allows approximately spherical flame balls to form. Their existence and stability are governed by the balance between chemical reaction, preferential diffusion, and heat loss \cite{Buckmaster1990, FernandezTarrazo2011}. Under terrestrial gravity, however, the combustion timescale is sufficiently long for buoyancy to become a leading-order effect. The hot kernel rises, generates a large-scale vortical flow, deforms into a cap-like structure, and may eventually split into several secondary kernels. Numerical simulations and large-scale experiments have shown that this convective motion can transport the flame over distances much greater than its initial size and at velocities well above the propagation rate expected for a conventional near-limit deflagration \cite{Yakovenko2018IJHE,Volodin2021CST,Yakovenko2021Fluids}.
	
	Recent studies provide further evidence that near-limit hydrogen flames cannot be described by a single classical propagation regime. Yanez et al.\ interpreted the breakup of continuous lean hydrogen flames in narrow gaps as the formation of isolated flame-ball-like structures maintained by preferential diffusion and strong heat loss \cite{Yanez2025Fire}. A later classification based on soliton theory showed that fragmented fronts may exhibit a much wider variety of morphologies and propagation modes than suggested by a simple distinction between continuous and discontinuous flames \cite{Yanez2026Fire}. Although wall confinement and heat losses make Hele--Shaw configurations different from freely rising kernels in large volumes, these studies support the broader conclusion that near-limit flame existence, morphology, and direction of motion emerge from coupled diffusive, thermal, and hydrodynamic processes rather than from mixture composition alone.
	
	A central aspect of all the discussed problems is the interaction between the flame and the flow that it generates. The buoyant rise of the hot products produces a toroidal vortex and an associated strain field. This flow may continuously deliver fresh mixture to some parts of the reaction zone, transport and rotate others, and impose strong axial and tangential stretch. Convection can therefore have both stabilizing and destabilizing effects, depending on the local flow topology. Combustion may persist along the upper and lateral surfaces of the kernel, while its lower or axial region is subjected to counterflow, convective cooling, and eventual local extinction. Classical analyses related the extinction of upward-propagating near-limit flames to stretch generated by buoyancy-driven motion \cite{Hertzberg1984}. More recent detailed calculations have shown that the vortex structure, axial strain, and inflow of cold fresh gas determine whether the lower reaction zone survives and whether the flame remains a connected cap or breaks into separate fragments \cite{Yakovenko2025Lebedev, Yakovenko2025PSEP}.
	
	A closely related question concerns the physical meaning of the flammability limit itself. Experimentally measured limits depend on the direction of flame propagation, vessel geometry, characteristic length scale, ignition source, wall heat losses, and the criterion used to define successful propagation. Convection is particularly important for upward-moving flames, whereas downward propagation requires the reaction zone to survive an adverse combination of flow and heat loss. Near the transition between these regimes, even small changes in mixture composition or flow conditions may produce major changes in flame shape, combustion completeness, and overpressure. Recent experiments and simulations have shown that the downward propagation limit cannot always be represented by a single concentration independent of ignition history and flow structure \cite{Anikin2022ICDERS, Yakovenko2025PSEP}. Similar effects arise when externally imposed or turbulence-generated velocity gradients alter residence times, local stretch, and species transport toward the reaction zone \cite{kiverin2024psaep}. Under practical terrestrial conditions, flammability limits should therefore be treated as outcomes of coupled chemistry, transport, heat loss, geometry, and flow, rather than as purely thermochemical properties of the fresh mixture.
	
	The discussed effects also complicate the definition and interpretation of burning velocity. For a steady planar premixed flame, the laminar burning velocity is an eigenvalue describing flame propagation relative to the unburned gas. Where such a planar solution exists, however, its eigenvalue needs not govern the evolution of a finite, highly curved, post-ignition kernel. Time-dependent calculations of centrally ignited lean hydrogen--air mixtures have identified flame-ball, radiatively extinguishing, and asymptotically planar-propagating branches in different composition ranges \cite{Tse2000PCI}. Flame-ball-based analyses have likewise emphasized that experimentally inferred or effective propagation velocities in very lean hydrogen mixtures may differ fundamentally from the one-dimensional planar value \cite{WilliamsGrcar2009,FernandezTarrazo2012IJHE}. In a buoyancy-dominated ultra-lean kernel, the observed motion additionally contains contributions from local flame displacement, translation of the hot-product region, radial and axial convection, and continuous deformation of the reaction surface. The leading-point velocity therefore cannot be identified directly with a laminar burning velocity. Similarly, lateral kernel growth may result from both deficient-reactant diffusion and vortex-induced spreading. Measurements of lean hydrogen flames have emphasized the difficulty of separating flame propagation from buoyant motion near the flammability limit \cite{Raman1998}. Large-scale experiments and flammability-tube studies likewise show that the apparent upward velocity may be controlled mainly by buoyant rise and, over part of the ultra-lean range, may depend only weakly on hydrogen concentration \cite{Volodin2021CST,Anikin2022ICDERS}.
	
	Among the molecular and thermal mechanisms governing near-limit combustion, radiative heat loss has received particular attention. Zel'dovich-type adiabatic flame balls are intrinsically unstable, whereas radiative heat loss can support stable flame-ball solutions in microgravity and determine both their radius and their lean existence limit \cite{Buckmaster1990}. Detailed studies of hydrogen flame balls have shown that radiation regulates the maximum temperature and establishes a balance between chemical heat release and energy loss \cite{FernandezTarrazo2011}. Its importance, however, depends strongly on the physical configuration. In burner-stabilized, ball-like flames under terrestrial gravity, convective losses and heat conduction to the burner may dominate over radiation \cite{HernandezPerez2015}. In narrow-gap flames, by contrast, fragmentation is governed mainly by intense heat transfer to the walls \cite{Yanez2025Fire}. The role of radiation in a freely rising, unconfined, buoyancy-deformed hydrogen flame kernel therefore cannot be inferred directly from either microgravity flame balls or wall-stabilized flames.
	
	Thermal diffusion, or the Soret effect, is the second physical mechanism of primary interest. Because molecular and atomic hydrogen are light and highly mobile, temperature gradients can generate appreciable thermal-diffusion fluxes. In lean hydrogen flames, these fluxes have a joint effect with preferential diffusion and flame curvature, altering the local mixture composition, radical concentrations, reaction rate, flame speed, and stability. Direct simulations of naturally propagating cellular flames showed that multicomponent transport and the Soret effect can produce hotter and substantially faster curved flames, together with qualitatively different extinction and cell-division behaviour \cite{Grcar2009}. Later studies confirmed that the influence of Soret diffusion depends strongly on flame stretch, pressure, temperature, curvature, and the stage of instability development \cite{Zhou2017}. Computationally efficient mixture-averaged formulations for thermal diffusion have also been validated against full multicomponent models for planar, stretched, steady, unsteady, laminar, and turbulent hydrogen flames \cite{Schlup2018CTM, Schlup2018CF}.
	
	Thermal diffusion becomes particularly important close to the lean flammability limit. Simulations of gravity-free hydrogen flame balls showed that Soret diffusion makes a non-negligible contribution to hydrogen transport toward the reaction zone and extends the flammability range \cite{FernandezTarrazo2011}. For burner-stabilized hydrogen--methane flames under terrestrial gravity, the inclusion of multicomponent transport together with Soret and Dufour effects increased the flame size, intensified fuel consumption, and shifted the stabilization position \cite{HernandezPerez2015}. These studies, however, considered either spherical structures without gravity or stationary flames controlled by an imposed flow and burner geometry.
	
	Despite the extensive literature on flame balls, lean hydrogen-flame instability, thermal diffusion, radiative stabilization, and buoyancy-driven propagation, the combined influence of Soret diffusion and radiative heat loss on a freely developing ultra-lean flame kernel under terrestrial gravity has not yet been quantified systematically. Earlier terrestrial-gravity studies established the general sequence of kernel rise, vortex formation, cap development, and breakup, but typically neglected thermal diffusion, radiation, or both \cite{Yakovenko2018IJHE, Volodin2021CST, Yakovenko2021Fluids, Yakovenko2025Lebedev, makhviladze1982cesw, kopylov1983cesw, makhviladze1992cesw}. Conversely, investigations that treated radiation and Soret transport in detail focused mainly on microgravity flame balls, confined flames, burner-stabilized configurations, or cellular fronts without natural convection \cite{FernandezTarrazo2011, HernandezPerez2015, Grcar2009, Zhou2017}. It therefore remains unclear how these mechanisms affect the initial diffusion-controlled growth, the subsequent buoyant acceleration, flame morphology, the extent of the active reaction zone, and the eventual breakup of an unconfined kernel.
	
	The present study addresses this gap through detailed numerical simulations of the combustion development in a 6 vol.\% hydrogen--air mixture under terrestrial gravity. A complete two-factor matrix is considered, with Soret diffusion and radiative heat loss independently enabled and disabled. One-dimensional spherically symmetric simulations are first used to characterize the intrinsic kernel-growth regime and its sensitivity to both mechanisms. The later buoyancy-driven evolution is then examined in two-dimensional planar and axisymmetric configurations. The analysis considers flame kernel motion, flow topology, temperature and species distributions, elliptic-cap deformation, and breakup. The numerically predicted leading-point trajectories are also compared with available experimental measurements. The aim is to determine whether radiation and thermal diffusion mainly alter the timescale of kernel evolution or whether they also change its morphological and stability pathways, thereby providing a quantitative link between molecular transport, heat loss, flame--vortex interaction, and hydrogen-fire hazards under terrestrial conditions.

	%%%%%%%%%%%%%%%%%%%%%%%%%%%%%%%%%%%%%%%%%%
	\section{Problem Setup}
	\label{sec:problem_setup}

	We considered a quiescent, premixed hydrogen--air mixture containing 6 vol.\% H$_2$ at an initial temperature of $T_0=300$~K and pressure of $p_0=101325$~Pa. Dry air was represented by an O$_2$:N$_2$ molar ratio of $1:3.762$, and the initial velocity was zero. Each geometry was simulated with four combinations of physical models: neither Soret diffusion nor radiation, Soret diffusion only, radiation only, and both effects together. This simulations set separates the individual effects of thermal diffusion and radiative heat loss from their combined effect.

\begin{figure}[htbp]
		\centering
		\includegraphics[width=0.75\linewidth]{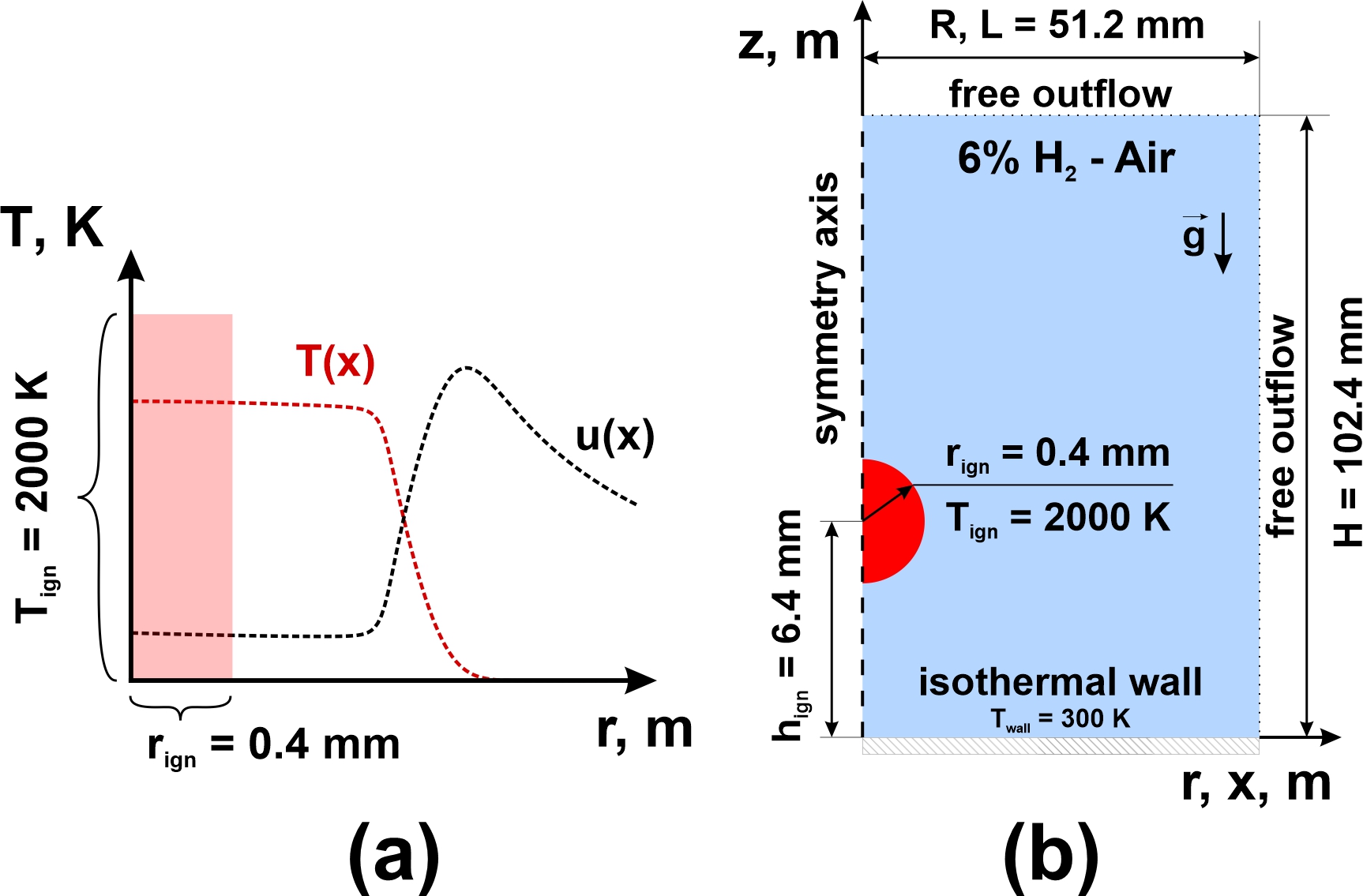}
		\caption{Computational configurations. (\textbf{a}) One-dimensional spherical reference problem with a centrally heated ignition region of radius $r_{\mathrm{ign}}=0.4$~mm at $T_{\mathrm{ign}}=2000$~K. (\textbf{b}) Half-domain used in the planar and axisymmetric calculations. In the planar case, $x$ and $L$ are the transverse coordinate and half-width; in the axisymmetric case, $r$ and $R$ are the radial coordinate and domain radius. The ignition region is located at the symmetry axis $h_{\mathrm{ign}}=6.4$~mm above the isothermal lower wall.}
		\label{fig:problem_setup}
	\end{figure}

	We performed two groups of simulations (Figure~\ref{fig:problem_setup}). The one-dimensional calculations used spherical symmetry to describe post-ignition kernel growth without buoyancy-induced deformation. Ignition was imposed by instantaneously heating a central sphere of radius $r_{\mathrm{ign}}=0.4$~mm to $T_{\mathrm{ign}}=2000$~K while keeping the initial composition unchanged. We imposed symmetry at $r=0$ and free outflow at the outer boundary. 

	The buoyancy-driven stage was simulated in two-dimensional planar and axisymmetric geometries. In both cases, we resolved one half of the physical domain. In the planar formulation, $x$ is the transverse coordinate and $L=51.2$~mm is the half-width. In the axisymmetric formulation, $r$ is the radial coordinate and $R=51.2$~mm is the domain radius. The domain height was $H=102.4$~mm. Symmetry was imposed at $x=0$ or $r=0$. The upper and outer lateral boundaries were free outflows, while the lower boundary was a slip isothermal wall at $T_{\mathrm{wall}}=300$~K. Gravity, $g=9.8$~m~s$^{-2}$, acted vertically downward. Ignition was initiated by heating a circular region centred on the symmetry axis at $h_{\mathrm{ign}}=6.4$~mm above the wall. The ignition radius and temperature were $r_{\mathrm{ign}}=0.4$~mm and $T_{\mathrm{ign}}=2000$~K. Both domains used a uniform spacing of $0.4$~mm, giving $128\times256$ cells in the resolved half-domain. Previous grid-convergence studies showed that a cell size of 0.4~mm is sufficient to reproduce the structure of the	energy-release zone in ultra-lean hydrogen--air flames \cite{yakovenko2024aa, Yakovenko2021Fluids}. 

	%%%%%%%%%%%%%%%%%%%%%%%%%%%%%%%%%%%%%%%%%%
	\section{Mathematical Model}
	\label{sec:mathematical_model}

	\subsection{Governing Equations}

	The reacting gas was modelled as a thermally perfect ideal-gas mixture under the low-Mach-number approximation of the gas dynamics \cite{Baum1978}. Pressure was split into a spatially uniform thermodynamic part and a hydrodynamic perturbation,
	\begin{equation}
		p(\boldsymbol{x},t)=p_0(t)+\pi(\boldsymbol{x},t)
		\label{eq:pressure_decomposition}
	\end{equation}
	where $p_0=101325$~Pa remained constant. The model solves conservation equations for mass, species, momentum, and sensible enthalpy \cite{Kuo}:
	\begin{equation}
		\frac{\partial \rho}{\partial t}+\boldsymbol{\nabla}\cdot(\rho\boldsymbol{u})=0,
		\label{eq:continuity}
	\end{equation}
	\begin{equation}
		\frac{\partial (\rho Y_k)}{\partial t}
		+\boldsymbol{\nabla}\cdot(\rho Y_k\boldsymbol{u})
		=-\boldsymbol{\nabla}\cdot\boldsymbol{j}_k+\dot{\omega}_k,
		\qquad k=1,\ldots,N_s,
		\label{eq:species}
	\end{equation}
	\begin{equation}
		\frac{\partial \boldsymbol{u}}{\partial t}
		+(\boldsymbol{u}\cdot\boldsymbol{\nabla})\boldsymbol{u}
		=-\frac{1}{\rho}\boldsymbol{\nabla}\pi
		+\frac{1}{\rho}\boldsymbol{\nabla}\cdot\boldsymbol{\tau}
		+\frac{\rho-\rho_0}{\rho}\boldsymbol{g},
		\label{eq:momentum}
	\end{equation}
	\begin{align}
		\frac{\partial (\rho h_s)}{\partial t}
		+\boldsymbol{\nabla}\cdot(\rho h_s\boldsymbol{u})
		={}&\frac{\mathrm{d}p_0}{\mathrm{d}t}
		+\boldsymbol{\nabla}\cdot(\lambda\boldsymbol{\nabla}T)
		-\boldsymbol{\nabla}\cdot\left(\sum_{k=1}^{N_s}h_{s,k}\boldsymbol{j}_k\right)
		\nonumber\\
		&-\sum_{k=1}^{N_s}\Delta h_{f,k}^{0}\dot{\omega}_k
		+\boldsymbol{\tau}:\boldsymbol{\nabla}\boldsymbol{u}
		+\dot{q}_{\mathrm{rad}}.
		\label{eq:enthalpy}
	\end{align}
	Here, $\rho$ is the mixture density, $\boldsymbol{u}$ is the mass-averaged velocity, $Y_k$ is the mass fraction of species $k$, $\boldsymbol{j}_k$ is its diffusive mass flux, and $\dot{\omega}_k$ is its chemical production rate. The quantities $h_s$ and $h_{s,k}$ are the mixture and species sensible enthalpies, $\lambda$ is the mixture thermal conductivity, $\Delta h_{f,k}^{0}$ is the standard enthalpy of formation, and $\dot{q}_{\mathrm{rad}}$ is the volumetric radiative source. The reference density $\rho_0$ is the ambient-mixture density. It appears in Equation~\eqref{eq:momentum} because the hydrostatic pressure component has been removed. The viscous stress tensor is
	\begin{equation}
		\boldsymbol{\tau}=\mu\left[
		\boldsymbol{\nabla}\boldsymbol{u}
		+(\boldsymbol{\nabla}\boldsymbol{u})^{\mathsf{T}}
		-\frac{2}{3}(\boldsymbol{\nabla}\cdot\boldsymbol{u})\boldsymbol{I}
		\right],
		\label{eq:stress}
	\end{equation}
	where $\mu$ is the dynamic viscosity.

	The thermodynamic and caloric closure was based on a thermally perfect ideal-gas mixture. The equation of state was written as
	\begin{equation}
		p_0=\rho R_{\mathrm{mix}}T,
		\qquad
		R_{\mathrm{mix}}=\frac{R_u}{\overline{M}}
		=R_u\sum_{k=1}^{N_s}\frac{Y_k}{M_k},
		\qquad
		\overline{M}=\left(\sum_{k=1}^{N_s}\frac{Y_k}{M_k}\right)^{-1},
		\label{eq:eos}
	\end{equation}
	where $R_u$ is the universal gas constant, $M_k$ is the molar mass of species $k$, $R_{\mathrm{mix}}$ is the mixture-specific gas constant, and $\overline{M}$ is the mixture molar mass. The sensible enthalpy appearing in Equation~\eqref{eq:enthalpy} was evaluated from the temperature-dependent species heat capacities according to
	\begin{equation}
		h_s(T,\boldsymbol{Y})=\sum_{k=1}^{N_s}Y_k h_{s,k}(T),
		\qquad
		h_{s,k}(T)=\int_{T_{\mathrm{ref}}}^{T}c_{p,k}(T')\,\mathrm{d}T',
		\qquad
		c_p(T,\boldsymbol{Y})=\sum_{k=1}^{N_s}Y_k c_{p,k}(T),
		\label{eq:sensible_enthalpy}
	\end{equation}
	where $T_{\mathrm{ref}}$ is the reference temperature associated with the standard-state thermochemical data. With this convention, the total species enthalpy is $h_k=\Delta h_{f,k}^{0}+h_{s,k}$, which leads to the chemical source $-\sum_k \Delta h_{f,k}^{0}\dot{\omega}_k$ in the sensible-enthalpy equation. Temperature-dependent heat capacities and enthalpies were represented by polynomial fits to JANAF thermochemical data \cite{Chase1998JANAF}. Hydrogen oxidation followed the detailed mechanism of K\'eromn\`es et al.~\cite{Keromnes2013}, with nitrogen treated as an inert gas.

	\subsection{Molecular Transport and Thermal Diffusion}

	Binary diffusion coefficients, species viscosities, and species thermal conductivities were calculated from kinetic theory using Lennard--Jones parameters and collision integrals. Mixture-averaged diffusion coefficients followed the zeroth-order Hirschfelder--Curtiss approximation \cite{Hirschfelder1954}. Mixture viscosity and thermal conductivity were evaluated with arithmetic--harmonic mean expressions,
	\begin{equation}
		\mu=\frac{1}{2}\left[
		\sum_{k=1}^{N_s}X_k\mu_k
		+\left(\sum_{k=1}^{N_s}\frac{X_k}{\mu_k}\right)^{-1}
		\right],
		\qquad
		\lambda=\frac{1}{2}\left[
		\sum_{k=1}^{N_s}X_k\lambda_k
		+\left(\sum_{k=1}^{N_s}\frac{X_k}{\lambda_k}\right)^{-1}
		\right],
		\label{eq:mixture_transport}
	\end{equation}
	where $X_k$, $\mu_k$, and $\lambda_k$ are the mole fraction, viscosity, and thermal conductivity of species $k$.

	The uncorrected mixture-averaged diffusion flux contains concentration and thermal-diffusion terms,
	\begin{equation}
		\boldsymbol{j}^{(0)}_k
		=-\rho D_k\boldsymbol{\nabla}Y_k
		-D_k^{T}\boldsymbol{\nabla}\ln T.
		\label{eq:raw_diffusion_flux}
	\end{equation}
	A correction velocity enforces zero total diffusive mass flux:
	\begin{equation}
		\boldsymbol{j}_k
		=\boldsymbol{j}^{(0)}_k
		-Y_k\sum_{l=1}^{N_s}\boldsymbol{j}^{(0)}_l,
		\qquad
		\sum_{k=1}^{N_s}\boldsymbol{j}_k=\boldsymbol{0}.
		\label{eq:corrected_flux}
	\end{equation}
	When the Soret effect was disabled, all $D_k^T$ were set to zero. When it was enabled, thermal diffusion of H and H$_2$ was described by the reduced Schlup--Blanquart model \cite{Schlup2018CF,Schlup2018CTM}. The coefficients were $\alpha_{\mathrm{H}}=0.895$ and $\alpha_{\mathrm{H_2}}=0.910$. Restricting thermal diffusion to H and H$_2$ follows the reduced-model formulation: in lean hydrogen flames, thermal diffusion primarily modifies the transport and profiles of these light species, whereas its direct influence on heavier-species profiles is much weaker; the H/H$_2$ reduced model reproduces multicomponent thermal-diffusion results with good accuracy \cite{Schlup2018CF,Schlup2018CTM}. The reciprocal Dufour contribution to the heat flux was neglected because, in premixed flames, opposing reactant and product concentration gradients make this cross-diffusion term comparatively small. For freely propagating lean H$_2$--air flames, Soret transport has been identified as the more relevant cross-diffusion contribution \cite{Grcar2009}. Sensible-enthalpy transport by both Fickian and Soret species fluxes was nevertheless retained in Equation~\eqref{eq:enthalpy}.

	\subsection{Radiative Heat Loss}

	Radiative cooling was described with an optically thin emission model. Water vapour was the only radiating species, and the volumetric source was
	\begin{equation}
		\dot{q}_{\mathrm{rad}}
		=-4\sigma_{\mathrm{SB}}\left(T^4-T_b^4\right)\kappa_P,
		\qquad T_b=300~\mathrm{K},
		\label{eq:radiation_source}
	\end{equation}
	where $\sigma_{\mathrm{SB}}$ is the Stefan--Boltzmann constant and $T_b$ is the background radiation temperature. The Planck-mean absorption coefficient was calculated as
	\begin{equation}
		\kappa_P
		=X_{\mathrm{H_2O}}\frac{p_0}{p_{\mathrm{ref}}}
		\sum_{n=0}^{5}a_n\left(\frac{1000}{T}\right)^n,
		\qquad p_{\mathrm{ref}}=101325~\mathrm{Pa},
		\label{eq:planck_mean}
	\end{equation}
	using the polynomial correlation of Barlow et al.~\cite{Barlow2001}.

	The optically thin approximation is suitable for these centimetre-scale kernels because they are much smaller than the absorption length of lean hydrogen-flame products \cite{FernandezTarrazo2011}.

	\subsection{Numerical Method}

	Equations~\eqref{eq:continuity}--\eqref{eq:enthalpy} were solved with the in-house NRG implementation of the low-Mach FDS algorithm described in \cite{McGrattan} and validated for solving hydrogen combustion problems in \cite{Yakovenko2023Fire}. The predictor--corrector projection scheme advanced density and species together with chemical and transport source terms. The equation of state then provided the required velocity divergence. A variable-coefficient Poisson equation supplied the hydrodynamic pressure before velocity correction and was solved with a multigrid method. No turbulence or subgrid combustion model was used because the flow remained strictly laminar and calculations directly resolved the flame structure.

\section{Results and Discussion}
\label{sec:results_discussion}

\subsection{Initial Diffusion-Controlled Growth in Spherical Geometry}
\label{sec:spherical_growth}

The one-dimensional spherical case isolates the early post-ignition growth of the flame kernel before natural convection changes its shape. This stage sets the initial conditions for the later buoyant motion. It determines both the volume of hot, low-density products on which buoyancy acts and the combustion intensity produced by molecular transport, chemistry, and heat loss. The spherical results therefore provide a reference for interpreting the multidimensional simulations.

\begin{figure}[htbp]
    \centering
    \includegraphics[width=0.80\linewidth]{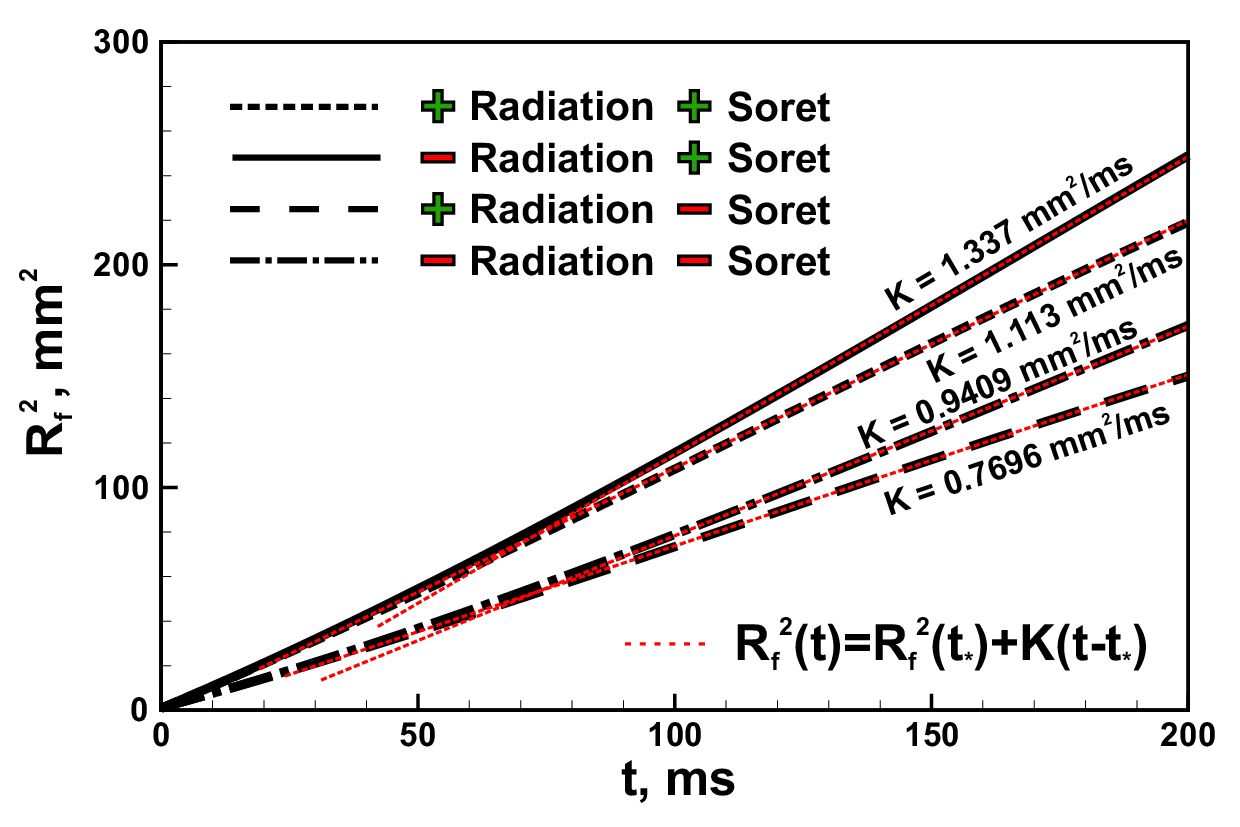}
    \caption{Squared flame-front radius, $R_f^2$, in the one-dimensional spherical simulation of a 6~vol.\% H$_2$--air mixture. The four curves represent the combinations of Soret diffusion and radiative heat loss shown in the legend. The front is defined by the maximum temperature gradient. Red dotted lines are fits of $R_f^2(t)=R_f^2(t_*)+K(t-t_*)$. The nearly linear curves indicate diffusion-controlled growth rather than constant-speed spherical expansion. Soret diffusion increases $K$, while radiation decreases it.}
    \label{fig:spherical_dynamics}
\end{figure}

Figure~\ref{fig:spherical_dynamics} shows the squared flame-front radius for all four combinations of Soret diffusion and radiative heat loss. After the ignition transient stage, the fitted part of each curve is well described by
\begin{equation}
    R_f^2(t)=R_f^2(t_*)+K(t-t_*),
    \label{eq:spherical_growth_law}
\end{equation}
where $t_*$ is the start of the fitting interval and $K$ is an effective kernel-growth coefficient. The main result is that $R_f^2$ has an almost constant slope. Differentiating Equation~\eqref{eq:spherical_growth_law} gives
\begin{equation}
    U_f=\frac{\mathrm{d}R_f}{\mathrm{d}t}=\frac{K}{2R_f}.
    \label{eq:spherical_front_velocity}
\end{equation}
The tracked front velocity thus decreases approximately as $R_f^{-1}$ rather than remaining constant. By contrast, a spherical deflagration approaching a constant burned-gas propagation velocity $S_b$ would follow $R_f\simeq R_{f,0}+S_b t$, and the slope of $R_f^2(t)$ would increase with time.

This result does not mean that a planar laminar burning velocity cannot be defined for the same nominal mixture under other conditions. It means that a constant planar reference speed does not describe the finite, strongly curved kernel studied here. This distinction matters in ultra-lean hydrogen combustion because different configurations can follow different solution branches. Tse et al.\ showed that centrally ignited hydrogen--air mixtures may form flame balls, extinguish through radiation, or approach planar propagation depending on composition \cite{Tse2000PCI}. In their model, planar propagation appeared only at hydrogen concentrations well above the flame-ball range. Other near-limit studies have described propagating cellular flames as assemblies of flame balls and have distinguished their effective propagation velocity from the burning velocity of a homogeneous planar flame \cite{WilliamsGrcar2009,FernandezTarrazo2011,FernandezTarrazo2012IJHE}. The exact concentration thresholds depend on the model, but these studies show that near-limit kernel motion need not be governed by a single planar-flame velocity.

A simple transport scaling helps explain Equation~\eqref{eq:spherical_growth_law}. If the deficient reactant diffuses toward the reaction zone over a distance of order $R_f$, the total diffusive supply scales as
\begin{equation}
    \dot m_D
    \sim 4\pi R_f^2\rho D_{\mathrm{eff}}\frac{\Delta Y}{R_f}
    \propto R_f,
    \label{eq:spherical_diffusive_supply}
\end{equation}
where $D_{\mathrm{eff}}$ is an effective transport scale rather than a single binary diffusivity. The mass consumption required to expand the spherical front scales as
\begin{equation}
    \dot m_F\sim4\pi R_f^2\rho_u\dot R_f.
    \label{eq:spherical_front_consumption}
\end{equation}
Balancing these rates gives $\dot R_f\propto R_f^{-1}$ and therefore $\mathrm{d}R_f^2/\mathrm{d}t\simeq\mathrm{const}$. This scaling does not predict the numerical value of $K$. Chemistry is still essential because it maintains the thermal and concentration gradients \cite{tereza2023rjpcba, tereza2025hfacp}. The coefficient $K$ includes the combined effects of preferential diffusion, heat conduction, finite-rate chemistry, thermal expansion, Soret transport, and radiative loss. It should be viewed as an effective kernel-growth coefficient, not as a molecular diffusivity or laminar burning velocity.

This growth law also controls the volume of hot gas available for buoyant acceleration,
\begin{equation}
    V_h=\frac{4}{3}\pi R_f^3.
    \label{eq:spherical_hot_volume}
\end{equation}
When the intercept in Equation~\eqref{eq:spherical_growth_law} is small compared with $Kt$,
\begin{equation}
    V_h\sim\frac{4\pi}{3}(Kt)^{3/2}.
    \label{eq:spherical_volume_scaling}
\end{equation}
A moderate change in $K$ can therefore produce a larger relative change in hot-product volume. The integrated buoyancy force can be written as
\begin{equation}
    F_b=g\int_{\Omega_h}(\rho_u-\rho)\,\mathrm{d}V
    \approx g(\rho_u-\overline{\rho}_h)V_h,
    \label{eq:integrated_buoyancy}
\end{equation}
where $\rho_u$ is the unburned-mixture density and $\overline{\rho}_h$ is the mean density of the hot products. Thus, the transport and heat-loss processes that set $K$ during the spherical stage also affect the initial buoyant force.

The curve ordering in Figure~\ref{fig:spherical_dynamics} shows that thermal diffusion has the strongest effect on the spherical growth rate. The Soret effect increases $K$ both with and without radiation. This increase is linked to thermodiffusive transport of H and H$_2$ toward the hot reaction zone. Hydrogen is the deficient reactant in this ultra-lean mixture, so a larger hydrogen flux strengthens local chemical conversion and allows the front to advance faster. Earlier studies also found strong Soret effects in curved and stretched lean hydrogen flames, where thermal diffusion changes local enrichment, reaction rates, and flame motion \cite{Grcar2009,Zhou2017,Schlup2018CTM}.

Radiation has the opposite effect and lowers $K$ in both Soret cases. It removes part of the released chemical energy, cools the products, and weakens the thermal and concentration gradients that support outward growth. The spherical calculations therefore show two competing trends: Soret diffusion accelerates the transport-controlled expansion, while radiation slows it. Under the present conditions, the Soret effect on kernel radius is clearly stronger.

\begin{figure}[htbp]
    \centering
    \includegraphics[width=\linewidth]{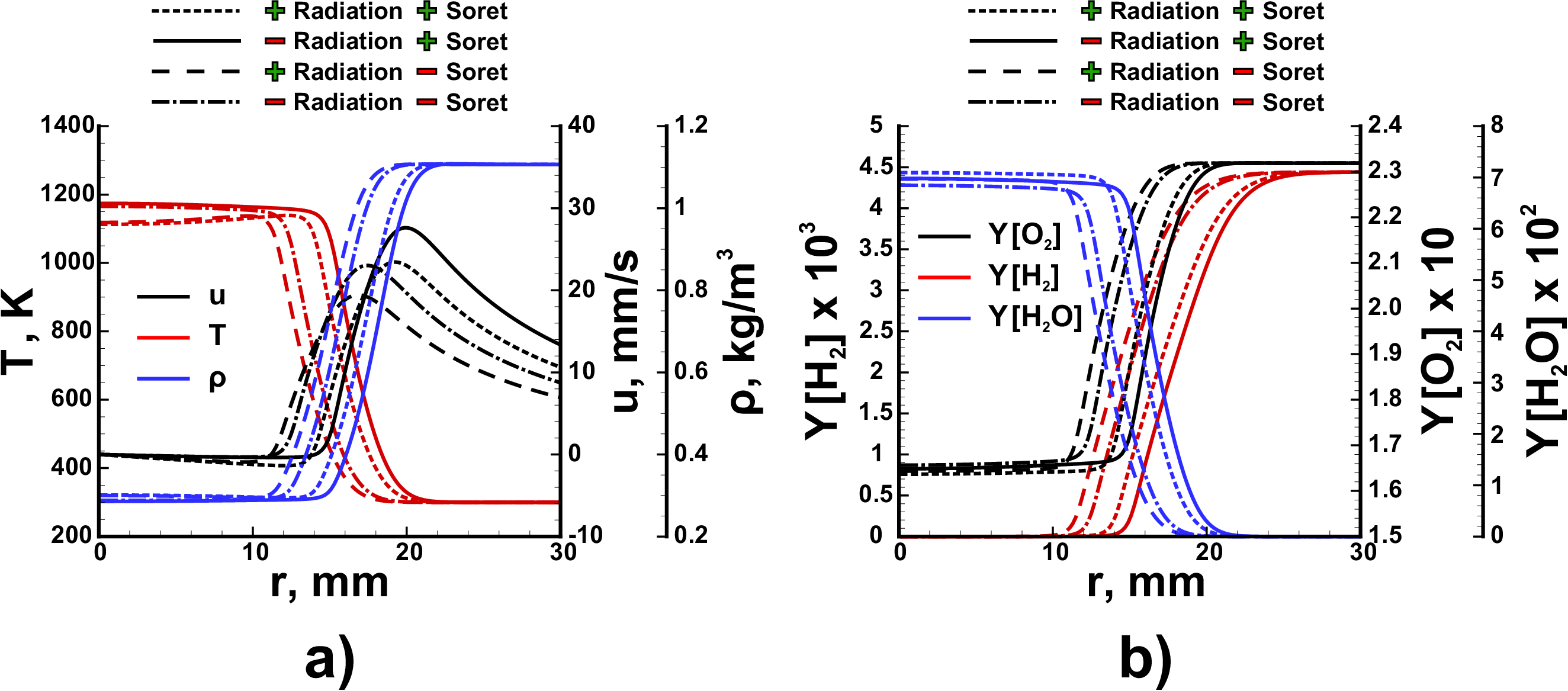}
    \caption{Radial structure of the spherical flame kernel at $t=200$~ms for all four model combinations. (\textbf{a}) Temperature $T$, radial velocity $u_r$, and density $\rho$ profiles. (\textbf{b}) Mass fractions of H$_2$, O$_2$, and H$_2$O profiles. Line styles identify the model combinations shown in the legend.}
    \label{fig:spherical_structure}
\end{figure}

The radial profiles in Figure~\ref{fig:spherical_structure} help explain these trends. At $t=200$~ms, radiation lowers the product temperature by about 61~K, which is roughly 7\% of the temperature rise above ambient. The product density increases by about 0.015~kg~m$^{-3}$, or only 1.89\% of the ambient-mixture density. The density response is therefore much smaller than the temperature change alone might suggest.

At the nearly uniform thermodynamic pressure of the low-Mach model, density is given by
\begin{equation}
    \rho=\frac{p_0\overline{M}}{R_uT},
    \label{eq:density_temperature_composition}
\end{equation}
and its relative change can be approximated by
\begin{equation}
    \frac{\delta\rho}{\rho}\approx
    \frac{\delta\overline{M}}{\overline{M}}-\frac{\delta T}{T}.
    \label{eq:density_variation}
\end{equation}
The small density increase indicates that the effect of cooling is partly offset by a change in the mean molar mass. The species profiles in Figure~\ref{fig:spherical_structure}b show small composition differences between the radiating and non-radiating cases. Radiation also produces a weak inward velocity in the product region. Without gravity or multidimensional recirculation, this motion results from thermal contraction of the cooled products and the radial-flow adjustment required by mass conservation. Radiation can therefore weaken later buoyant motion in two ways: by reducing the hot-product volume through a smaller $K$ and by reducing the density deficit.

The Soret effect changes the flame structure in a different way. It moves the reaction zone outward and increases the kernel radius, but it hardly changes the plateau temperature or density of the fully burned products. Its main signature appears in the species profiles. Thermal diffusion increases hydrogen transport into the hot zone, which leads to greater oxygen consumption and water production. Soret diffusion therefore mainly changes the reactant supply and the rate at which new products are formed, rather than the thermodynamic state of products that have already formed.

This distinction matters for the multidimensional stage. In the Soret-enabled cases, the larger buoyant force at the onset of rise comes mainly from the larger volume of hot products, not from a larger density difference per unit volume. Radiation reduces both the volume and the density difference. The spherical simulations thus explain the later differences in rise velocity and flame deformation.

The transition from nearly spherical growth to buoyant motion can be estimated with the timescale criterion of Leblanc et al.~\cite{Leblanc2013PoF}. Their approach compares the time needed for a flame to grow over a distance of order $R$ with the time needed for buoyancy to move the kernel over the same distance. The rise velocity is estimated from the Davies--Taylor relation for spherical bubbles,
\begin{equation}
    V_{\mathrm{rise}}=\frac{2}{3}\sqrt{g'R},
    \qquad
    g'=g\left(1-\frac{\rho_b}{\rho_u}\right),
    \label{eq:leblanc_rise_velocity}
\end{equation}
where $\rho_b$ and $\rho_u$ are representative burned- and unburned-gas densities. Leblanc et al.\ assumed an approximately constant spherical growth velocity. For the present kernel, the radius-dependent velocity in Equation~\eqref{eq:spherical_front_velocity} gives
\begin{equation}
    \Pi_{bK}
    =\frac{t_{\mathrm{growth}}}{t_{\mathrm{rise}}}
    =\frac{V_{\mathrm{rise}}}{\dot R_f}
    =\frac{4}{3}\frac{\sqrt{g'R_f^3}}{K}.
    \label{eq:buoyancy_diffusive_growth}
\end{equation}
Values of $\Pi_{bK}\ll1$ correspond to diffusion-controlled, nearly isotropic growth, while $\Pi_{bK}=O(1)$ marks the expected onset of noticeable buoyant translation and deformation. The corresponding radius is
\begin{equation}
    R_{bK}^{*}
    =\left(\frac{9K^2}{16g'}\right)^{1/3}.
    \label{eq:critical_buoyancy_radius}
\end{equation}
For the case with both Soret diffusion and radiation, $K=1.113\times10^{-3}$~m$^2$~s$^{-1}$ and representative densities of $0.3011$ and $1.1066$~kg~m$^{-3}$ give an order-of-magnitude estimate  $R_{bK}^{*}\approx4.6$~mm. 

\subsection{Buoyancy-Driven Flame Evolution and Experimental Validation}
\label{sec:buoyant_dynamics}

The multidimensional simulations show how the transport and heat-loss effects identified in the spherical case influence the later buoyant motion. Soon after ignition, the flame remains nearly spherical and follows the diffusion-controlled growth described in Section~\ref{sec:spherical_growth}. As the kernel grows and the ratio in Equation~\eqref{eq:buoyancy_diffusive_growth} approaches unity, buoyancy drives it upward and creates a vortical flow. This flow gradually deforms the kernel into a cap.

\begin{figure}[htbp]
    \centering
    \includegraphics[width=\linewidth]{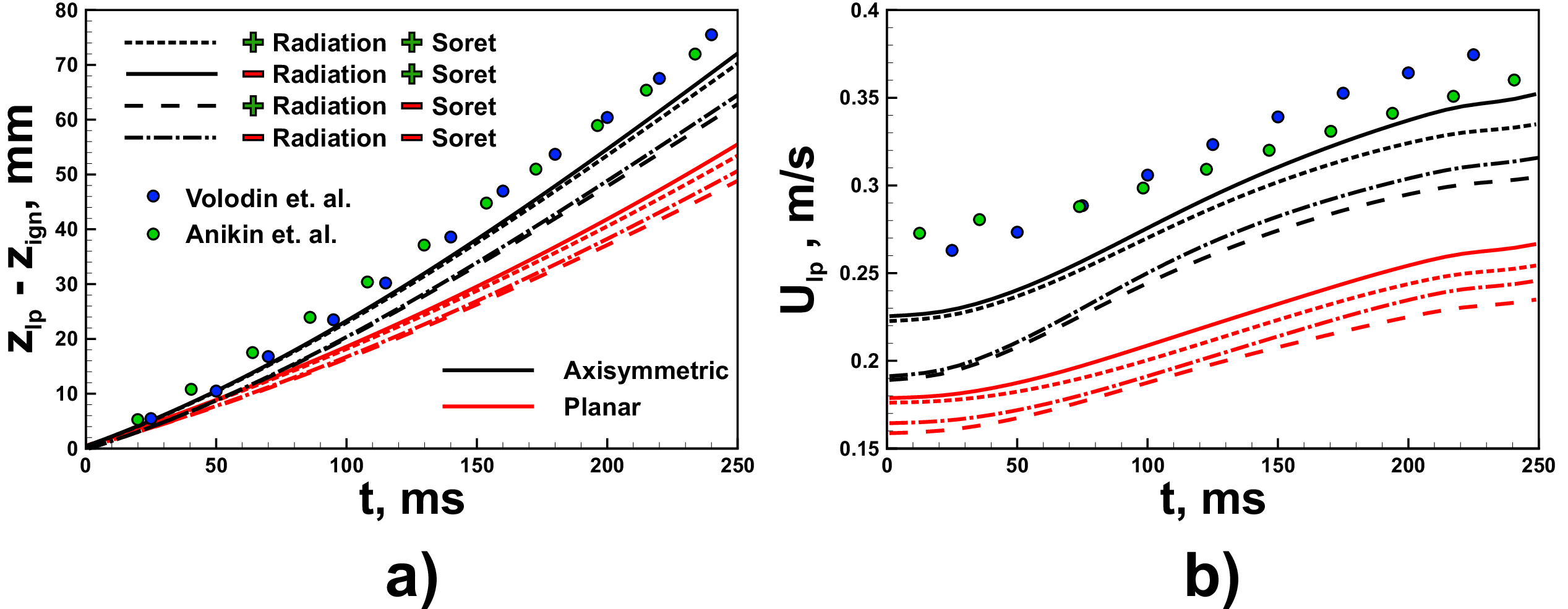}
    \caption{Leading-point displacement from the ignition position, $z_{lp}-z_{\mathrm{ign}}$, and leading-point velocity, $U_{lp}$, in the planar and axisymmetric simulations of a 6~vol.\% H$_2$--air flame kernel. The curves show the four combinations of Soret diffusion and radiation. Data from Volodin et al.~\cite{Volodin2021CST} and from Anikin et al.~\cite{Anikin2022ICDERS} are used for the validation.}
    \label{fig:flame_dynamics}
\end{figure}

Figure~\ref{fig:flame_dynamics} compares the leading-point trajectories and velocities from the planar and axisymmetric simulations with experimental data. Both geometries reproduce the accelerating rise and give the same ordering of the four model combinations. The planar simulations, however, consistently underpredict the displacement. The axisymmetric results agree much more closely with the experiment. This confirms that the surrounding-flow geometry strongly affects buoyant kernel motion, as earlier comparisons between planar simulations and experiments had suggested \cite{Volodin2021CST}.

The experimental flame is a finite three-dimensional object, so the axisymmetric model is the better approximation for this comparison. The discussion below therefore focuses mainly on the axisymmetric results. The planar results are retained as a measure of geometry sensitivity. Separating the dimensionality effect into buoyancy, added mass, drag, and shape contributions would require fully three-dimensional simulations and is outside the scope of this study.

In both geometries, Soret diffusion increases the leading-point displacement, while radiation causes a smaller reduction. The same ordering was found in the spherical simulations. Thermal diffusion increases the rate at which hot products are formed and thus strengthens the later buoyant motion. Radiation lowers both the growth rate and the temperature-dependent density deficit. Differences created during the early transport-controlled stage are therefore carried into, and amplified during, the buoyant stage.

\begin{figure}[htbp]
    \centering
    \includegraphics[width=\linewidth]{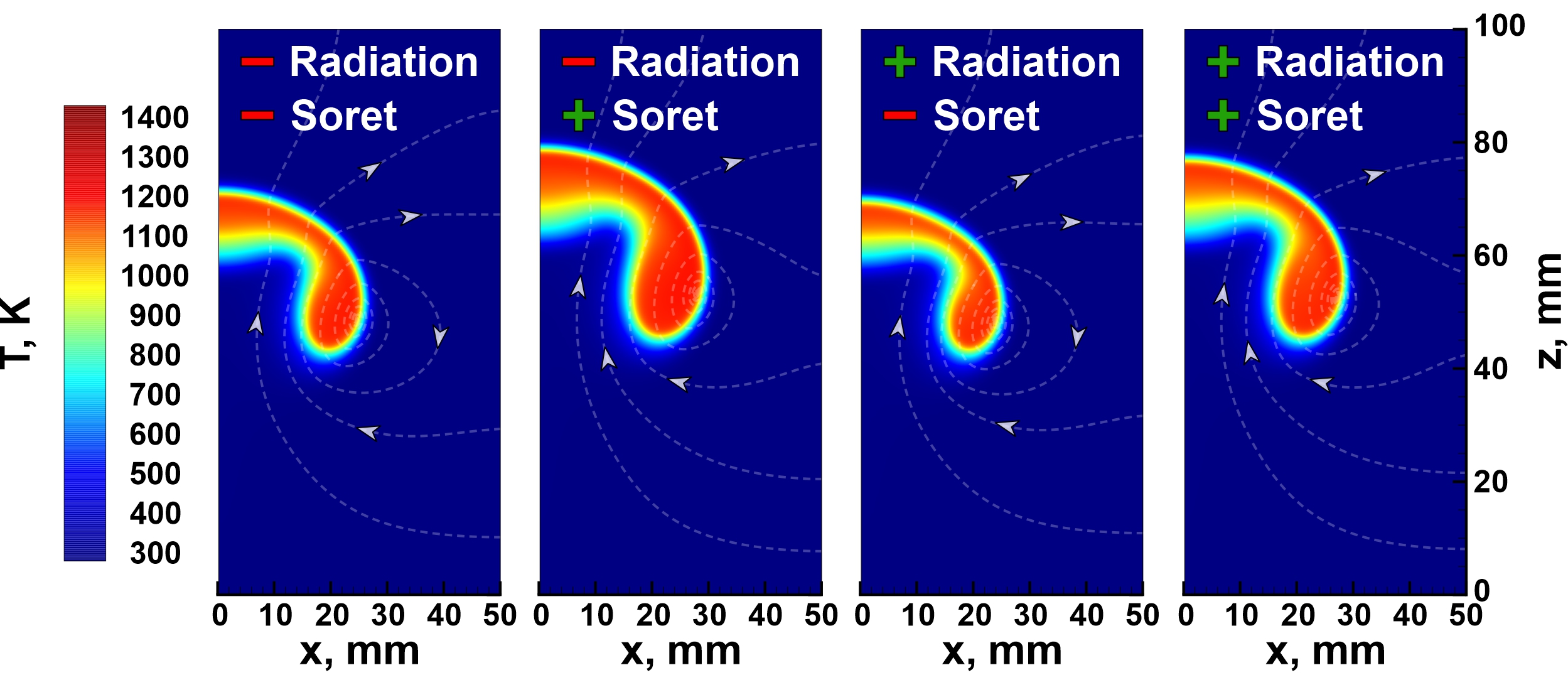}
    \caption{Temperature fields in the axisymmetric simulation at $t=250$~ms for a 6~vol.\% H$_2$--air mixture. From left to right: no radiation/no Soret diffusion, no radiation/with Soret diffusion, with radiation/no Soret diffusion, and both effects enabled. All panels use the same temperature scale.}
    \label{fig:axi_temperature}
\end{figure}

Figure~\ref{fig:axi_temperature} shows the axisymmetric temperature fields at $t=250$~ms. In every case, the initially compact kernel has developed into a cap-shaped hot region with a clear wake. The upper boundary moves upward and spreads laterally, while the lower boundary becomes deeply indented. These features are produced by the toroidal vortex generated as the hot products rise. The flow draws fresh mixture toward the lower part of the kernel and carries the lateral flame sections around the vortex core. This creates the cap shape reported in earlier simulations and experiments \cite{Yakovenko2018IJHE,Volodin2021CST,Yakovenko2021Fluids}.

The four model combinations retain the same relative behaviour seen in the spherical case. Kernels with Soret diffusion are larger and extend farther upward and sideways. Radiation slightly lowers the product temperature and reduces the kernel size. The small difference between the radiating and non-radiating cases confirms that radiation plays a secondary role in the early and intermediate rise under these conditions. It may still become more important near breakup, where modest changes in temperature and reaction rate can determine whether a local flame segment survives.

\begin{figure}[htbp]
    \centering
    \includegraphics[width=\linewidth]{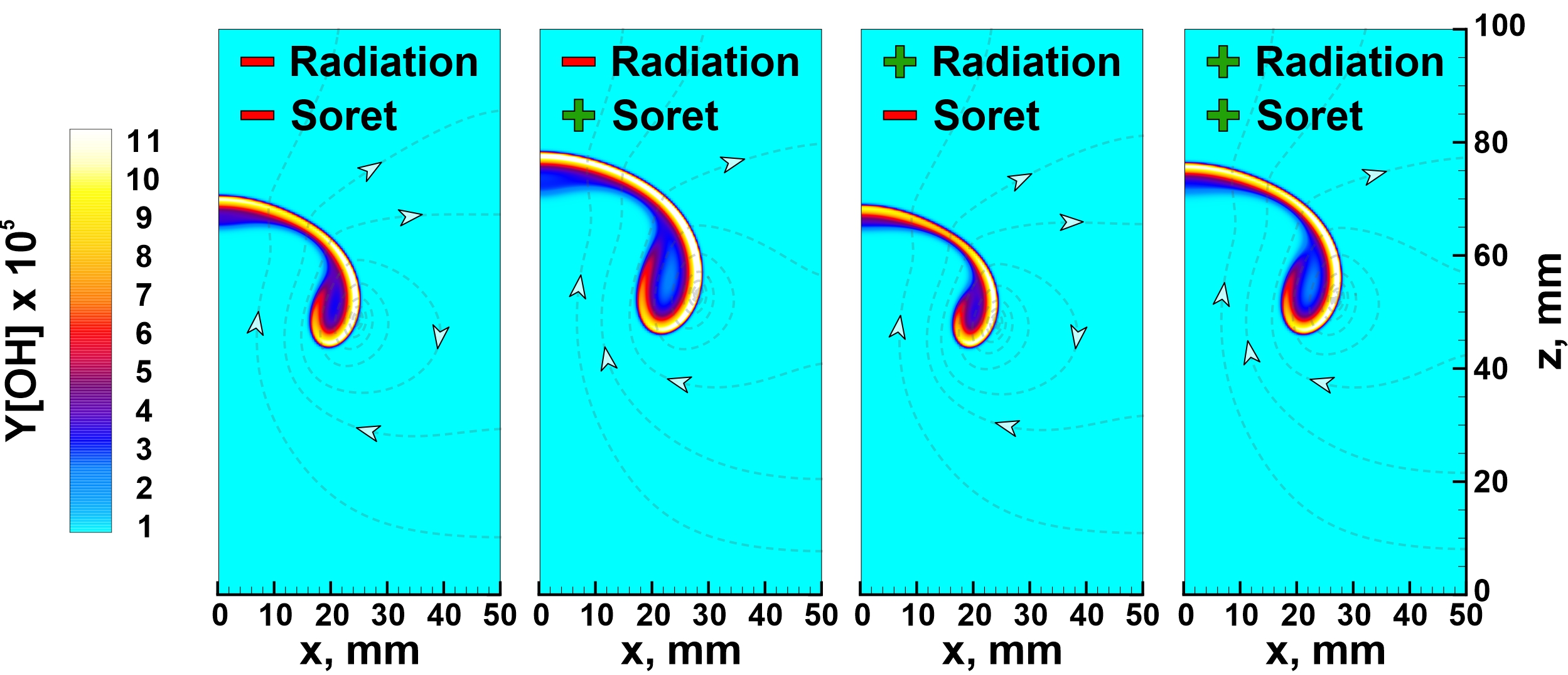}
    \caption{OH mass-fraction fields in the axisymmetric simulation at $t=250$~ms. The case order is the same as in Figure~\ref{fig:axi_temperature}. All panels use the same OH scale.}
    \label{fig:axi_oh}
\end{figure}

The OH fields in Figure~\ref{fig:axi_oh} separate the active reaction zone from the wider hot-product region. The temperature field forms a large cap and wake, but appreciable OH is confined mainly to the upper surface and lateral shoulders. Much of the lower thermal boundary has little chemical activity. The rising kernel is therefore not a closed flame propagating uniformly in all directions.

The toroidal vortex continuously draws colder unburned gas toward the lower part of the kernel. This axial inflow cools and strains the flame near the axis, which weakens or extinguishes the local reaction zone. Earlier studies of buoyant ultra-lean hydrogen flames found the same pattern: an active upper cap above an open lower region transported mainly as hot products \cite{Yakovenko2018IJHE,Yakovenko2021Fluids,Yakovenko2025Lebedev}. The flame--vortex interaction thus creates the cap shape and prepares the kernel for breakup.

Soret diffusion enlarges and displaces the OH-rich layer. This agrees with the stronger hydrogen transport and greater oxygen consumption found in the spherical profiles. Radiation slightly shrinks the active region but does not change its location on the upper side of the kernel. The front topology is therefore set mainly by preferential transport and self-induced convection, while radiation changes its extent only modestly.

The axisymmetric trajectories agree well with the measurements of Volodin et al.~\cite{Volodin2021CST} and Anikin et al.~\cite{Anikin2022ICDERS} and remove the systematic underprediction of the planar model. Both datasets show that the observed upward motion is governed mainly by buoyant translation and deformation of the hot kernel, not by a conventional laminar burning velocity.

These results connect the early spherical growth to the later buoyant evolution. Soret diffusion increases the production of hot gas, giving a larger kernel and faster rise. Radiation lowers the product temperature, density deficit, and growth rate, but has a smaller effect on the trajectory. Once buoyancy dominates, the self-induced vortex reorganizes the flame into an active upper cap above a mostly inactive lower wake. This structure sets the stage for the deformation and breakup discussed next.

\subsection{Evolution of the Flame-Kernel Morphology}
\label{sec:flame_morphology}

Buoyancy changes both the position and the shape of the flame kernel. At first, the kernel remains nearly spherical, in agreement with the one-dimensional results considered in Section~\ref{sec:spherical_growth}. As the hot-product volume grows and the kernel begins to rise, it becomes an oblate cap. The upper boundary expands upward and sideways, while the lower boundary is reshaped by the wake vortex and the inflow of cold gas along the axis. Similar cap-shaped and fragmented flames have been observed in earlier experiments and simulations of ultra-lean hydrogen mixtures \cite{Yakovenko2018IJHE,Volodin2021CST,Yakovenko2021Fluids}.

To measure this deformation, we fitted the upper part of the flame with an axis-aligned ellipse,
\begin{equation}
	\frac{(x-x_c)^2}{a_r^2}
	+
	\frac{(z-z_c)^2}{a_z^2}
	=
	1,
	\label{eq:elliptic_cap}
\end{equation}
where $a_r$ and $a_z$ are the lateral and vertical semi-axes. In the axisymmetric case, $x$ is replaced by the radial coordinate $r$. The ellipse represents only the upper cap. It does not describe the full hot-product region, because the lower boundary becomes strongly distorted and partly open as the kernel develops.

\begin{figure}[H]
	\centering
	\includegraphics[width=0.8\textwidth]{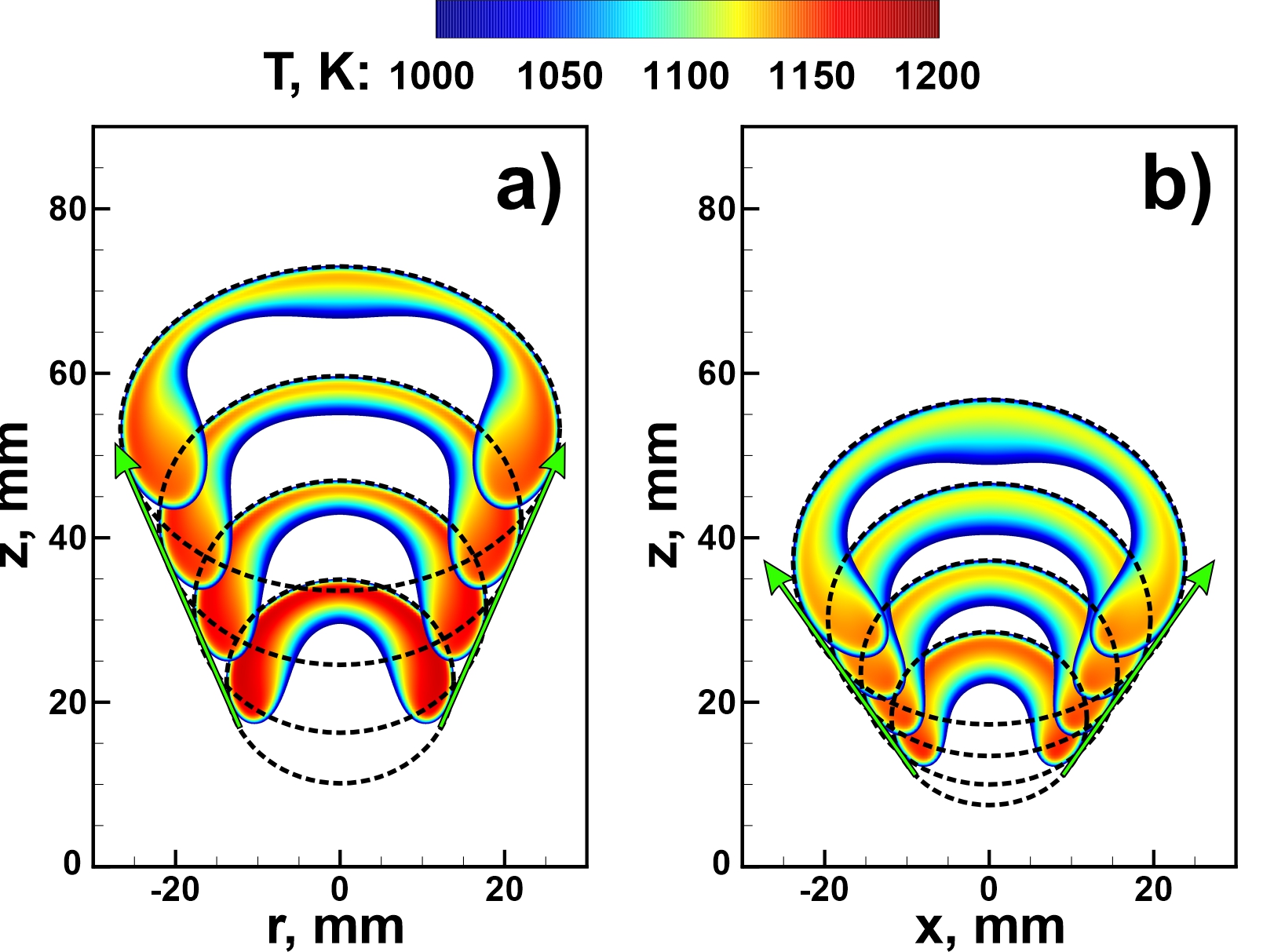}
	\caption{Flame-kernel shape in (\textbf{a}) axisymmetric and (\textbf{b}) planar simulations at successive stages of the rise. Black dashed curves are elliptic fits to the upper cap, and green lines show their approximately common lateral tangents.}
	\label{fig:flame_topology}
\end{figure}

Figure~\ref{fig:flame_topology} shows the flame-shape evolution in the axisymmetric and planar cases. In both geometries, the ellipses fit the upper boundary well through much of the rise. The fitted ellipses also share an approximately common lateral tangent. The cap therefore expands along a persistent geometric envelope rather than by isotropic radial growth. The tangent angle is smaller in the planar case, which means that the planar kernel spreads farther sideways for a given vertical displacement.

The elliptic shape shows where a spherical flame--bubble analogy breaks down. A spherical model is useful during the early diffusion-controlled stage and gives a simple estimate for the onset of buoyancy. Later, the growing difference between $a_r$ and $a_z$ makes the relevant hydrodynamic length scales direction-dependent. Acceleration, added mass, wake development, and drag should then depend on the changing aspect ratio. A fuller description would require an elliptic-bubble model for the planar case and an ellipsoidal-bubble model for the axisymmetric case. Developing such a model is beyond the scope of this work, but the systematic deformation provides a clear reason to pursue it.

\begin{figure}[H]
	\centering
	\includegraphics[width=0.8\textwidth]{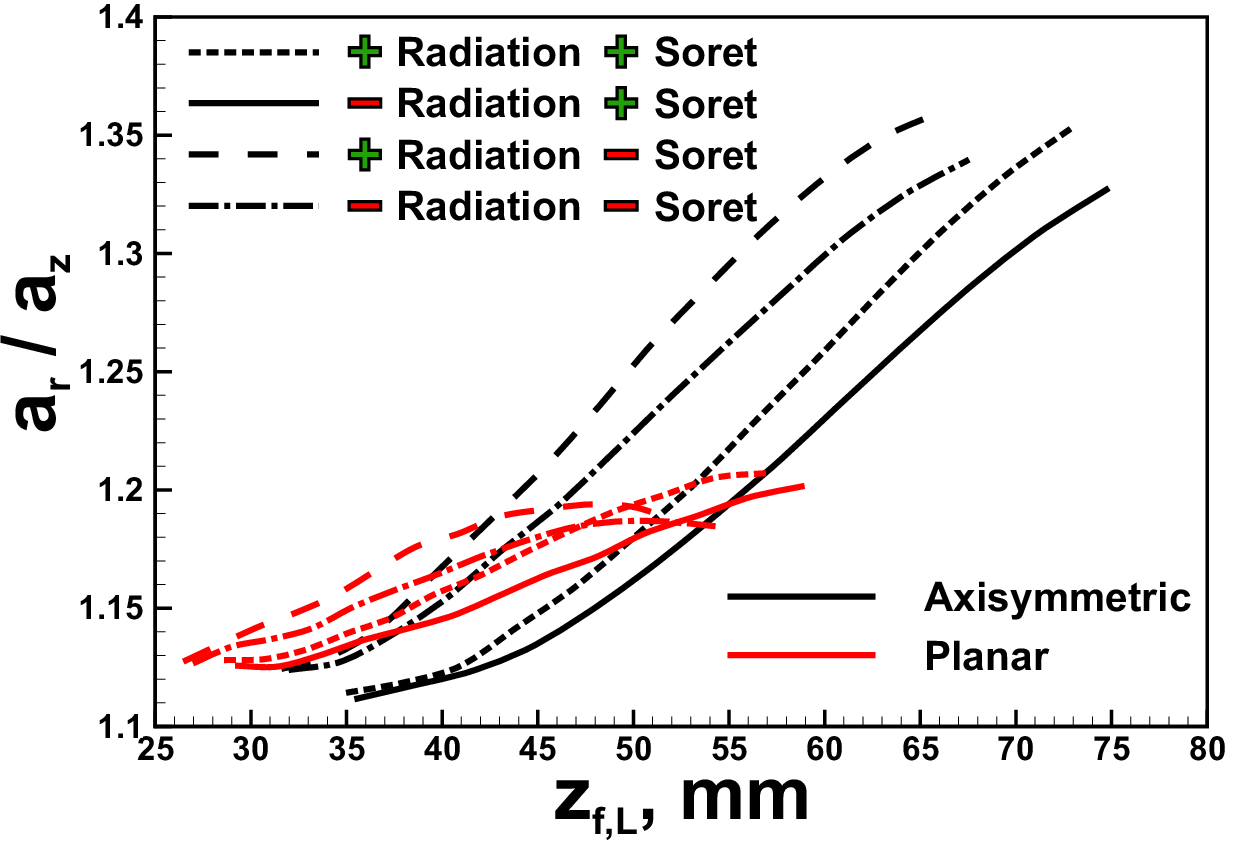}
	\caption{Aspect ratio of the fitted cap, $a_r/a_z$, versus leading-point coordinate $z_{\mathrm{lp}}$ for the planar (red) and axisymmetric (black) simulations. Curves show the four combinations of Soret diffusion and radiative heat loss.}
	\label{fig:shape_dynamics}
\end{figure}

The cap deformation is measured in Figure~\ref{fig:shape_dynamics} by the aspect ratio
\begin{equation}
	\chi
	=
	\frac{a_r}{a_z}.
	\label{eq:aspect_ratio}
\end{equation}
A value of $\chi=1$ represents a circular cap in the meridional plane, while $\chi>1$ represents an oblate cap that is wider than it is tall. We plot $\chi$ against the leading-point coordinate $z_{\mathrm{lp}}$ rather than time. This compares kernels at similar stages of vertical development and reduces the direct effect of their different rise speeds.

After an initial adjustment, $a_r/a_z$ generally rises with $z_{\mathrm{lp}}$ in both geometries. The cap therefore becomes progressively flatter as it moves upward, and the spherical approximation becomes less accurate. Formally,
\begin{equation}
	\frac{\mathrm{d}}{\mathrm{d}t}
	\ln\left(\frac{a_r}{a_z}\right)
	=
	\frac{\dot a_r}{a_r}
	-
	\frac{\dot a_z}{a_z},
	\label{eq:aspect_ratio_rate}
\end{equation}
so an increasing aspect ratio means that the relative lateral growth rate exceeds the relative vertical growth rate.

The aspect ratio gives a global measure of the balance between vortex-driven lateral spreading and combustion-defined propagation of the upper cap. The lateral semi-axis $a_r$ is strongly influenced by the buoyant vortex, which carries the flame shoulders outward. The vertical semi-axis $a_z$ also reflects propagation of the active upper flame relative to the hot-product centroid. Thus, $a_r/a_z$ compares lateral deformation with vertical flame kernel growth. It is not a direct measure of local burning velocity or vortex strength because both axes include coupled effects of translation, reaction, convection, and deformation.

The geometry comparison depends on whether the kernels are compared at the same time or at the same leading-point position. At a fixed time, the axisymmetric kernel rises faster and reaches a larger $z_{\mathrm{lp}}$ (Figure~\ref{fig:flame_dynamics}). Over the common range of $z_{\mathrm{lp}}$, however, the planar curves in Figure~\ref{fig:shape_dynamics} generally lie above the axisymmetric curves. The planar kernel is therefore more strongly flattened at a similar vertical position. This agrees with the smaller tangent angle in Figure~\ref{fig:flame_topology}. The aspect ratio and the tangent angle independently show stronger relative lateral spreading in the planar model.

The main purpose of Figure~\ref{fig:shape_dynamics} is to compare the physical models. At the same $z_{\mathrm{lp}}$, radiation generally increases $a_r/a_z$ in both geometries, making the upper cap more oblate. Cooling lowers the product temperature and weakens the active upper flame, which limits vertical growth. Radiation also reduces buoyancy and weakens the vortex, but the simulations show that vertical cap growth is reduced more than lateral spreading. The net result is a larger aspect ratio.

Soret diffusion has the opposite effect on shape. It increases the overall kernel size and displacement but generally lowers $a_r/a_z$ at the same $z_{\mathrm{lp}}$. The upper cap therefore remains closer to circular. As shown by the spherical calculations, thermal diffusion carries more hydrogen into the hot reaction zone, increases oxygen consumption, and promotes product formation. In the multidimensional flame, this stronger reactant supply supports the active upper surface and its vertical growth, partly offsetting the lateral flattening imposed by the vortex.

Radiation and Soret diffusion therefore change the morphology in opposite directions. Radiation favours relative lateral spreading, while thermal diffusion supports more balanced cap growth. This differs from their effect on the leading-point trajectory, where Soret diffusion accelerates the rise and radiation slows it only slightly. The leading-point position alone is thus not enough to describe the model effects. Kernels at similar heights may have different aspect ratios, curvatures, and distributions of chemical activity.

The increasing aspect ratio also affects flame stability. A flatter cap changes the curvature and stretch of the active upper surface and alters its interaction with the wake vortex. It also increases the lateral distance over which the reaction zone must remain continuous. The shape formed during the rise therefore prepares the necking and breakup described in the next section.

\subsection{Flame--Vortex Interaction and Kernel Breakup}
\label{sec:flame_breakup}

\begin{figure}[h]
	\centering
	\includegraphics[width=0.9\linewidth]{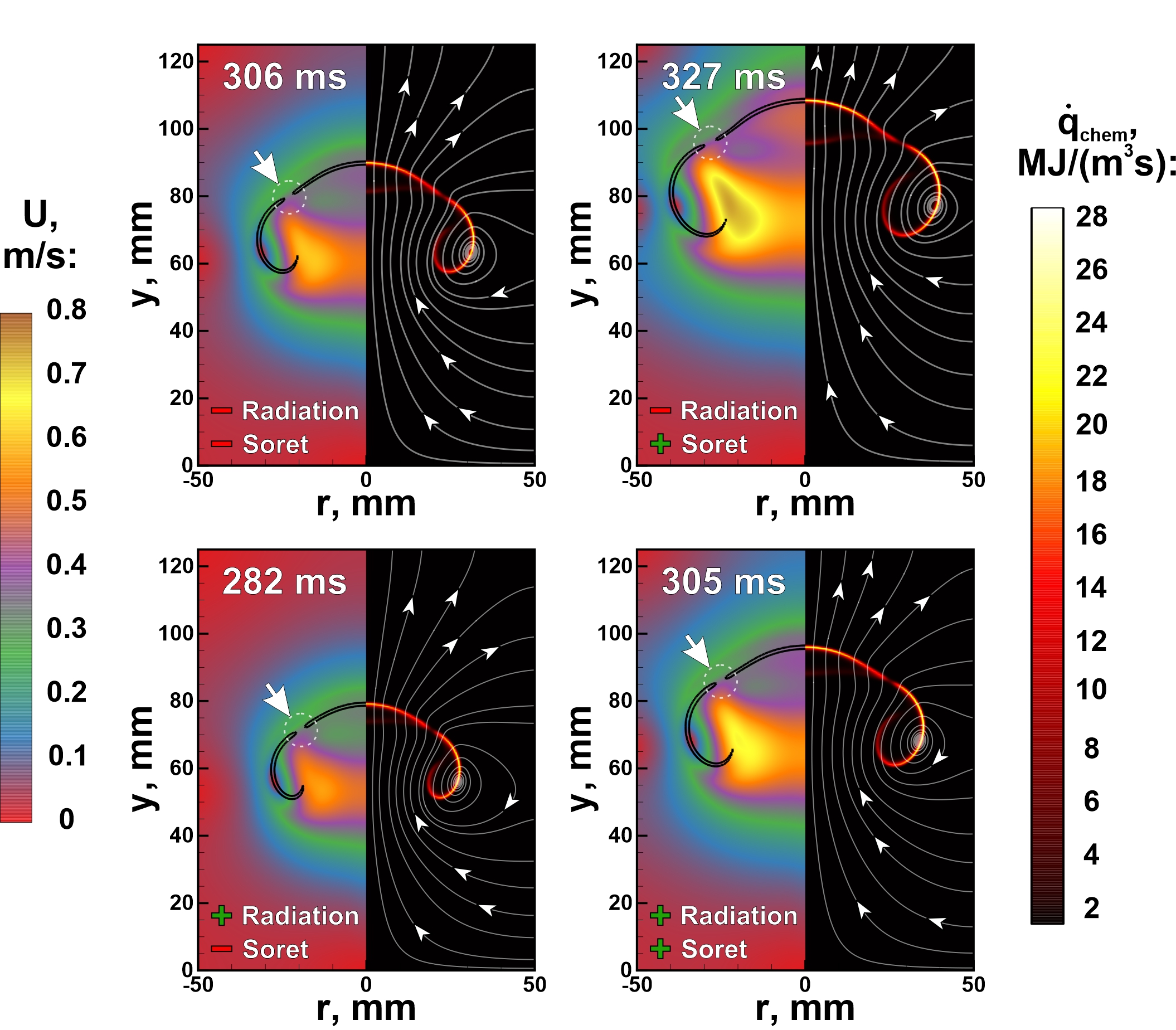}
	\caption{Axisymmetric flame structures just after breakup: (\textbf{a}) radiation off, Soret diffusion off, $t=306$~ms; (\textbf{b}) radiation off, Soret diffusion on, $t=327$~ms; (\textbf{c}) radiation on, Soret diffusion off, $t=282$~ms; and (\textbf{d}) radiation on, Soret diffusion on, $t=305$~ms. The left half of each panel shows velocity magnitude $U$ and the chemical heat release rate isolines $\dot{q}_{\mathrm{chem}}=10$~MJ~m$^{-3}$~s$^{-1}$. The right half shows streamlines and volumetric chemical heat-release rate  along the flame front. Large white arrows mark the new gap.}
	\label{fig:flame_breakup}
\end{figure}

The deformation described in Section~\ref{sec:flame_morphology} eventually breaks the continuity of the reaction zone. Figure~\ref{fig:flame_breakup} shows the axisymmetric flame just after the first interruption of the front for each model combination. The panels combine velocity magnitude, streamlines, and volumetric chemical heat-release rate along the flame front. Together, they show how the buoyancy-generated vortex causes breakup.

The flow pattern is similar in all four cases. Rising hot products create a toroidal vortex below and partly inside the flame cap. Fresh gas is drawn toward the kernel, turns around the vortex core, and moves outward along the upper structure. At the same time, the lower flame branch is rolled around the vortex. The upper cap and the curled lower branch then follow different paths and remain linked by a narrow, strongly deformed flame segment.

Breakup occurs in this connecting segment. Here, the vortex creates strong tangential strain and rapidly changes the flame curvature. The reaction zone is stretched, the residence time of reactants and radicals in the hot layer falls, and convective heat loss increases. When chemical heat release and preferential transport can no longer balance these losses, the reaction rate drops and a gap opens. The event is therefore a local stretch-induced extinction, not a simple mechanical tearing of an interface.

This interpretation agrees with earlier studies of terrestrial ultra-lean hydrogen flames. The rising kernel creates a toroidal vortex, axial flow, and a nonuniform strain field that determine which parts of the flame survive \cite{Yakovenko2021Fluids,Yakovenko2025Lebedev}. Large-scale experiments also showed that lateral spreading and stretch can cause local extinction, front separation, and the formation of secondary kernels \cite{Volodin2021CST}. The present simulations reproduce this sequence and show how Soret diffusion and radiation shift the breakup threshold.

The breakup times respond almost symmetrically to the two mechanisms. Without radiation, Soret diffusion delays breakup from $306$ to $327$~ms, a delay of $21$~ms. With radiation, it delays breakup from $282$ to $305$~ms, or $23$~ms. Under the present conditions, thermal diffusion extends the lifetime of the connected flame by about $7$--$8\%$.

This stabilizing effect follows from the earlier influence of Soret diffusion. Thermodiffusion increases the fluxes of H and H$_2$ into the hot reaction zone and promotes fuel consumption and H$_2$O formation. In the strongly curved and stretched segment that later breaks, this extra supply of reactants and radicals helps maintain chemical heat release despite the shorter residence time. Calculations of curved and stretched lean hydrogen flames have reported similar increases in local source terms and propagation rates when thermal diffusion is included \cite{Grcar2009,Zhou2017,Schlup2018CTM}. Soret diffusion does not remove the hydrodynamic instability, but it allows the flame to withstand the deformation for longer.

Radiation acts in the opposite direction. Without Soret diffusion, it advances breakup from $306$ to $282$~ms, a change of $24$~ms. With Soret diffusion, it advances breakup from $327$ to $305$~ms, a change of $22$~ms. Radiative cooling lowers the temperature and reaction rate and reduces the safety margin of the stretched flame segment. The connection therefore fails at an earlier stage of vortex development.

The two effects nearly cancel in terms of breakup time. With both mechanisms active, breakup occurs at $305$~ms, almost the same as the $306$~ms obtained when both are disabled. The two cases are not physically identical: their temperature, composition, size, and trajectory histories differ. Their opposing effects on local flame survival simply lead to nearly the same time of first front interruption. The small remaining difference also suggests that, for this observable, the interaction between the mechanisms is weaker than their separate effects.

Although the breakup times differ, the post-breakup structures are very similar. In every case, the cap separates near the region of strongest vortex interaction, and active flame remains on both sides of the gap. The models mainly change how long it takes to reach the critical local condition; they do not change the breakup path. Flow topology controls where breakup occurs, while transport and radiation control how well the flame resists the associated strain and cooling.

Breakup does not mean complete extinction. Chemical-energy release remains strong along the upper cap and the rolled lower branch after the front is interrupted. The original kernel becomes several reacting structures rather than disappearing. These fragments may move through the thermal and chemical wake, interact, merge, or form secondary kernels, as observed experimentally \cite{Volodin2021CST} and numerically \cite{Yakovenko2021Fluids}. A survival criterion based only on a flame front continuity would therefore be too restrictive for this near-limit regime.

The breakup also provides a useful link to turbulent flammability. The present axisymmetric flow is laminar and dominated by one coherent toroidal vortex, so it is not a turbulent flame. Even so, the local interaction contains the same basic processes found in flame--eddy interactions: front deformation, curvature and strain, changes in reactant residence time, enhanced convective transport, and local extinction when the flow timescale becomes shorter than the chemical and diffusive timescales.

The present case can thus be viewed as a controlled, large-scale example of a single flame--vortex interaction. In turbulence, many eddies of different sizes and strain rates create such interactions at the same time and at different locations. The turbulent flammability limit is therefore not set only by the existence of an adiabatic planar-flame solution. It also depends on whether enough of the flame surface can survive the local stretch, heat loss, and scalar dissipation imposed by the flow. Classical flame-stretch theory likewise links practical flammability and blowoff limits to the competition between flame propagation and flow-induced deformation \cite{Hertzberg1984}.

The results show that transport and thermal losses can shift a flow-controlled extinction threshold without changing the fresh-mixture composition. Soret diffusion helps the flame survive vortex-induced stretch, while radiation makes local extinction easier. In more complex flows, both mechanisms should affect local extinction probability and, in turn, practical turbulent flammability limits. A quantitative extension to turbulence would require non-axisymmetric simulations with a spectrum of interacting vortices and is beyond the scope of this work.

The breakup stage completes the sequence examined here. Soret diffusion and radiation first change the diffusion-controlled growth of the spherical kernel, then alter its buoyant motion and shape, and finally shift the time of vortex-induced local extinction and fragmentation. The breakup mechanism itself remains controlled by the flow, but the flame's resistance depends on the balance among preferential transport, chemical heat release, radiative cooling, and convective deformation.

\section{Conclusions}
\label{sec:conclusions}

The combined effects of Soret diffusion and radiative heat loss on the development of a buoyant 6~vol.\% H$_2$--air flame kernel were examined using 1D spherically symmetric, 2D planar, and axisymmetric simulations. During the initial gravity-independent stage, the flame radius followed $R_f^2\sim Kt$, indicating diffusion-controlled kernel growth rather than constant-speed spherical propagation. Soret diffusion increased the effective growth coefficient $K$, whereas radiation reduced it.

The axisymmetric calculations reproduced the measured leading-point trajectory substantially better than the planar formulation. Soret diffusion produced larger and faster-rising kernels, mainly by enhancing hydrogen transport toward the reaction zone and increasing the rate of hot-product formation. Radiation moderately reduced the kernel size, product temperature, density deficit, and rise rate.

Buoyancy progressively transformed the initially compact kernel into an elliptic cap with an inactive lower wake. Radiation promoted greater relative lateral flattening, while Soret diffusion maintained a more nearly circular upper cap. These results show that trajectory and morphology respond differently to molecular transport and heat loss and therefore cannot be characterized by a single propagation velocity.

In all model combinations, breakup resulted from interaction with the self-induced toroidal vortex. Vortex-induced strain caused local extinction in the narrow connection between the upper cap and the rolled lower branch. Soret diffusion delayed breakup, whereas radiation advanced it, although neither mechanism changed the qualitative fragmentation pathway. The breakup topology was therefore controlled primarily by the flow, while thermal diffusion and radiation determined the flame resistance to stretch and cooling.

Overall, near-limit hydrogen-flame development under terrestrial gravity is governed by the coupled effects of preferential transport, heat loss, buoyancy, and flame--vortex interaction. Fully three-dimensional simulations are required to extend these findings to non-axisymmetric breakup and turbulent flammability limits.

\authorcontributions{Conceptualization, I.S.Y.; methodology, I.S.Y.; software, I.S.Y.; validation, I.S.Y.; formal analysis, I.S.Y. and A.D.K; investigation, I.S.Y.; resources, I.S.Y. and A.D.K.; data curation, I.S.Y.; writing---original draft preparation, I.S.Y.; writing---review and editing, I.S.Y. and A.D.K.; visualization, I.S.Y.; supervision, I.S.Y. and A.D.K.; project administration, I.S.Y. and A.D.K.; funding acquisition, I.S.Y. and A.D.K. All authors have read and agreed to the published version of the manuscript.}

\funding{This work was supported by the Ministry of Science and Higher Education of the Russian Federation (State Assignment No. 075-00270-26-00).}

\conflictsofinterest{The authors declare no conflicts of interest.}

	%%%%%%%%%%%%%%%%%%%%%%%%%%%%%%%%%%%%%%%%%%
	\reftitle{References}
	\bibliography{references}

@article{Yakovenko2023Fire,
	author  = {Yakovenko, Ivan and Kiverin, Alexey},
	title   = {Numerical Modeling of Hydrogen Combustion: Approaches and Benchmarks},
	journal = {Fire},
	year    = {2023},
	volume  = {6},
	number  = {6},
	pages   = {239},
	doi     = {10.3390/fire6060239}
}

@article{Lin2024Fire,
	author  = {Lin, Yujie and Ling, Xiaodong and Yu, Anfeng and Liu, Yi and Liu, Di and Wang, Yazhen and Wu, Qian and Lu, Yuan},
	title   = {Modeling of Hydrogen Dispersion, Jet Fires and Explosions Caused by Hydrogen Pipeline Leakage},
	journal = {Fire},
	year    = {2024},
	volume  = {7},
	number  = {1},
	pages   = {8},
	doi     = {10.3390/fire7010008}
}

@article{Tian2025Fire,
	author  = {Tian, Yao and Zhang, Xia and Xia, Qing and Chen, Zhaoyang},
	title   = {Oxidation Mechanisms of Electrolyte and Fire Gas Generation Laws During a Lithium-Ion Battery Thermal Runaway},
	journal = {Fire},
	year    = {2025},
	volume  = {8},
	number  = {6},
	pages   = {226},
	doi     = {10.3390/fire8060226}
}

@article{Qiu2023Batteries,
	author  = {Qiu, Mingming and Liu, Jianghong and Cong, Beihua and Cui, Yan},
	title   = {Research Progress in Thermal Runaway Vent Gas Characteristics of Li-Ion Battery},
	journal = {Batteries},
	year    = {2023},
	volume  = {9},
	number  = {8},
	pages   = {411},
	doi     = {10.3390/batteries9080411}
}

@article{Yanez2025Fire,
	author  = {Yanez, Jorge and Kagan, Leonid and Kuznetsov, Mike and Sivashinsky, Gregory},
	title   = {On Disintegrating Lean Hydrogen Flames in Narrow Gaps},
	journal = {Fire},
	year    = {2025},
	volume  = {8},
	number  = {9},
	pages   = {345},
	doi     = {10.3390/fire8090345}
}

@article{Yanez2026Fire,
	author  = {Yanez, Jorge and Kuznetsov, Mike and Kagan, Leonid and Sivashinsky, Gregory},
	title   = {{2D} Flameballs: An Enhanced Classification Based on Soliton Theory},
	journal = {Fire},
	year    = {2026},
	volume  = {9},
	number  = {7},
	pages   = {288},
	doi     = {10.3390/fire9070288}
}

@article{Yakovenko2018IJHE,
	author  = {Yakovenko, I. S. and Ivanov, M. F. and Kiverin, A. D. and Melnikova, K. S.},
	title   = {Large-Scale Flame Structures in Ultra-Lean Hydrogen--Air Mixtures},
	journal = {International Journal of Hydrogen Energy},
	year    = {2018},
	volume  = {43},
	number  = {3},
	pages   = {1894--1901},
	doi     = {10.1016/j.ijhydene.2017.11.138}
}

@article{Volodin2021CST,
	author  = {Volodin, Vladislav V. and Golub, Victor V. and Kiverin, Alexey D. and Melnikova, Ksenia S. and Mikushkin, Anton Yu. and Yakovenko, Ivan S.},
	title   = {Large-Scale Dynamics of Ultra-Lean Hydrogen--Air Flame Kernels in Terrestrial Gravity Conditions},
	journal = {Combustion Science and Technology},
	year    = {2021},
	volume  = {193},
	number  = {2},
	pages   = {225--234},
	doi     = {10.1080/00102202.2020.1748606}
}

@article{Yakovenko2021Fluids,
	author  = {Yakovenko, Ivan and Kiverin, Alexey and Melnikova, Ksenia},
	title   = {Ultra-Lean Gaseous Flames in Terrestrial Gravity Conditions},
	journal = {Fluids},
	year    = {2021},
	volume  = {6},
	number  = {1},
	pages   = {21},
	doi     = {10.3390/fluids6010021}
}

@article{Yakovenko2025Lebedev,
	author  = {Yakovenko, I. S. and Kiverin, A. D. and Melnikova, K. S.},
	title   = {Effect of Convection on the Flame Stability and Dynamics in Ultralean Hydrogen--Air Mixtures},
	journal = {Bulletin of the Lebedev Physics Institute},
	year    = {2025},
	volume  = {52},
	number  = {Suppl. 2},
	pages   = {S121--S129},
	doi     = {10.3103/S1068335624602826}
}

@article{Yakovenko2025PSEP,
	author  = {Yakovenko, I. and Kiverin, A. and Melnikova, K. and Stakhanov, V. and Ryakin, A.},
	title   = {Downward Flame Propagation Mechanisms in Lean Hydrogen--Air Mixtures within Large Volumes},
	journal = {Process Safety and Environmental Protection},
	year    = {2025},
	volume  = {203},
	number  = {Part B},
	pages   = {108000},
	doi     = {10.1016/j.psep.2025.108000}
}

@inproceedings{Anikin2022ICDERS,
	author    = {Anikin, N. B. and Tyaktev, A. A. and Kirillov, I. A. and Simonenko, V. A.},
	title     = {Experimental Study of Early-Stage Dynamics of the Ascending and Descending Laminar Hydrogen--Air Flames in a Vertical Closed Rectangular Tube},
	booktitle = {28th International Colloquium on the Dynamics of Explosions and Reactive Systems (ICDERS)},
	address   = {Naples, Italy},
	year      = {2022},
	month     = {June},
	note      = {Paper 183}
}

@article{makhviladze1982cesw,
	title = {Combustion-Source Development in a Closed Vessel in Conditions of Natural Convection},
	author = {Makhviladze, G. M. and Nikolova, I. P.},
	year = {1982},
	month = sep,
	journal = {Combustion, Explosion, and Shock Waves},
	volume = {18},
	number = {5},
	pages = {530--536},
	publisher = {{Springer Science and Business Media LLC}},
	issn = {0010-5082, 1573-8345},
	doi = {10.1007/bf00800618},
	urldate = {2025-07-15},
	copyright = {http://www.springer.com/tdm},
	langid = {english},
}

@article{kopylov1983cesw,
	title = {Influence of Natural Convection on the Concentrational Limits of Ignition of a Fuel Mixture in a Closed Vessel},
	author = {Kopylov, G. G. and Makhviladze, G. M.},
	year = {1983},
	journal = {Combustion, Explosion, and Shock Waves},
	volume = {19},
	number = {2},
	pages = {135--141},
	publisher = {{Springer Science and Business Media LLC}},
	issn = {0010-5082, 1573-8345},
	doi = {10.1007/bf00789226},
	urldate = {2025-07-15},
	copyright = {http://www.springer.com/tdm},
	langid = {english},
}

@article{makhviladze1992cesw,
	title = {Flame Propagation in a Horizontal Channel with Cold Side Walls in a Gravitational Field},
	author = {Makhviladze, G. M. and Melikhov, V. I.},
	year = {1992},
	journal = {Combustion, Explosion, and Shock Waves},
	volume = {28},
	number = {1},
	pages = {18--24},
	publisher = {{Springer Science and Business Media LLC}},
	issn = {0010-5082, 1573-8345},
	doi = {10.1007/bf00754961},
	urldate = {2025-07-15},
	copyright = {http://www.springer.com/tdm},
	langid = {english},
}

@techreport{Hertzberg1984,
	author      = {Hertzberg, Martin},
	title       = {The Theory of Flammability Limits: Flow Gradient Effects and Flame Stretch},
	institution = {U.S. Bureau of Mines, United States Department of the Interior},
	year        = {1984},
	number      = {Report of Investigations 8865},
	address     = {Washington, DC, USA}
}

@techreport{Raman1998,
	author      = {Raman, Kumar S.},
	title       = {Laminar Burning Velocities of Lean Hydrogen--Air Mixtures},
	institution = {Graduate Aeronautical Laboratories, California Institute of Technology},
	year        = {1998},
	number      = {Explosion Dynamics Laboratory Report FM97-15},
	address     = {Pasadena, CA, USA}
}

@article{kiverin2024psaep,
	title = {Dynamic Loads Induced by Near-Limit Turbulent Hydrogen-Air Combustion inside a Confinement},
	author = {Kiverin, Alexey and Melnikova, Ksenia and Yakovenko, Ivan},
	date = {2024-09},
	journal = {Process Safety and Environmental Protection},
	shortjournal = {Process Saf. Environ. Prot.},
	volume = {189},
	pages = {728--735},
	issn = {09575820},
	doi = {10.1016/j.psep.2024.06.034},
	url = {https://linkinghub.elsevier.com/retrieve/pii/S0957582024007171},
	urldate = {2025-05-25},
	langid = {english}
}

@article{Buckmaster1990,
	author  = {Buckmaster, J. and Joulin, G. and Ronney, P. D.},
	title   = {The Structure and Stability of Nonadiabatic Flame Balls},
	journal = {Combustion and Flame},
	year    = {1990},
	volume  = {79},
	number  = {3--4},
	pages   = {381--392},
	doi     = {10.1016/0010-2180(90)90147-J}
}

@article{FernandezTarrazo2011,
	author  = {Fern{\'a}ndez-Tarrazo, E. and S{\'a}nchez, A. L. and Li{\~n}{\'a}n, A. and Williams, F. A.},
	title   = {The Structure of Lean Hydrogen--Air Flame Balls},
	journal = {Proceedings of the Combustion Institute},
	year    = {2011},
	volume  = {33},
	number  = {1},
	pages   = {1203--1210},
	doi     = {10.1016/j.proci.2010.05.086}
}

@article{HernandezPerez2015,
	author  = {Hern{\'a}ndez-P{\'e}rez, Francisco E. and Oostenrijk, Bart and Shoshin, Yuriy and van Oijen, Jeroen A. and de Goey, Laurentius P. H.},
	title   = {Formation, Prediction and Analysis of Stationary and Stable Ball-Like Flames at Ultra-Lean and Normal-Gravity Conditions},
	journal = {Combustion and Flame},
	year    = {2015},
	volume  = {162},
	number  = {3},
	pages   = {932--943},
	doi     = {10.1016/j.combustflame.2014.09.020}
}

@article{Grcar2009,
	author  = {Grcar, Joseph F. and Bell, John B. and Day, Marcus S.},
	title   = {The Soret Effect in Naturally Propagating, Premixed, Lean, Hydrogen--Air Flames},
	journal = {Proceedings of the Combustion Institute},
	year    = {2009},
	volume  = {32},
	number  = {1},
	pages   = {1173--1180},
	doi     = {10.1016/j.proci.2008.06.075}
}

@article{Zhou2017,
	author  = {Zhou, Zhen and Hern{\'a}ndez-P{\'e}rez, Francisco E. and Shoshin, Yuriy and van Oijen, Jeroen A. and de Goey, Laurentius P. H.},
	title   = {Effect of Soret Diffusion on Lean Hydrogen/Air Flames at Normal and Elevated Pressure and Temperature},
	journal = {Combustion Theory and Modelling},
	year    = {2017},
	volume  = {21},
	number  = {5},
	pages   = {879--896},
	doi     = {10.1080/13647830.2017.1311028}
}

@article{Schlup2018CTM,
	author  = {Schlup, Jason and Blanquart, Guillaume},
	title   = {Validation of a Mixture-Averaged Thermal Diffusion Model for Premixed Lean Hydrogen Flames},
	journal = {Combustion Theory and Modelling},
	year    = {2018},
	volume  = {22},
	number  = {2},
	pages   = {264--290},
	doi     = {10.1080/13647830.2017.1398350}
}

@article{Schlup2018CF,
	author  = {Schlup, Jason and Blanquart, Guillaume},
	title   = {A Reduced Thermal Diffusion Model for {H} and {H2}},
	journal = {Combustion and Flame},
	year    = {2018},
	volume  = {191},
	pages   = {1--8},
	doi     = {10.1016/j.combustflame.2017.12.022}
}

@article{Alvarez2024Ammonia,
	author  = {Alvarez, Luis F. and Shaffer, James and Dumitrescu, Cosmin E. and Askari, Omid},
	title   = {Laminar Burning Velocity of Ammonia/Air Mixtures at High Pressures},
	journal = {Fuel},
	year    = {2024},
	volume  = {363},
	pages   = {130986},
	doi     = {10.1016/j.fuel.2024.130986}
}

@article{babkin1982cesw,
	title = {Effect of Tube Diameter on Homogeneous Gas Flame Propagation Limits},
	author = {Babkin, V. S. and Zamashchikov, V. V. and Badalyan, A. M. and Krivulin, V. N. and Kudryavtsev, E. A. and Baratov, A. N.},
	date = {1982},
	journal = {Combustion, Explosion, and Shock Waves},
	shortjournal = {Combust. Explos. Shock Waves},
	volume = {18},
	number = {2},
	pages = {164--171},
	issn = {0010-5082, 1573-8345},
	doi = {10.1007/BF00789613},
	url = {http://link.springer.com/10.1007/BF00789613},
	urldate = {2024-10-12},
	langid = {english}
}

@article{Keromnes2013,
	author  = {K{\'e}romn{\`e}s, Alan and Metcalfe, Wayne K. and Heufer, Karl A. and Donohoe, Nicola and Das, Apurba K. and Sung, Chih-Jen and Herzler, J{\"u}rgen and Naumann, Clemens and Griebel, Peter and Mathieu, Olivier and Krejci, Michael C. and Petersen, Eric L. and Pitz, William J. and Curran, Henry J.},
	title   = {An Experimental and Detailed Chemical Kinetic Modeling Study of Hydrogen and Syngas Mixture Oxidation at Elevated Pressures},
	journal = {Combustion and Flame},
	year    = {2013},
	volume  = {160},
	number  = {6},
	pages   = {995--1011},
	doi     = {10.1016/j.combustflame.2013.01.001}
}

@book{Chase1998JANAF,
	author    = {Chase, Malcolm W., Jr.},
	title     = {{NIST-JANAF} Thermochemical Tables},
	edition   = {4},
	series    = {Journal of Physical and Chemical Reference Data Monograph},
	volume    = {9},
	publisher = {American Institute of Physics},
	address   = {Woodbury, NY, USA},
	year      = {1998},
	isbn      = {9781563968310},
	doi       = {10.18434/T42S31}
}

@book{Hirschfelder1954,
	author    = {Hirschfelder, Joseph O. and Curtiss, Charles F. and Bird, R. Byron},
	title     = {Molecular Theory of Gases and Liquids},
	publisher = {John Wiley \& Sons},
	address   = {New York, NY, USA},
	year      = {1954},
	isbn      = {9780471400653}
}

@article{Barlow2001,
	author  = {Barlow, R. S. and Karpetis, A. N. and Frank, J. H. and Chen, J.-Y.},
	title   = {Scalar Profiles and {NO} Formation in Laminar Opposed-Flow Partially Premixed Methane/Air Flames},
	journal = {Combustion and Flame},
	year    = {2001},
	volume  = {127},
	number  = {3},
	pages   = {2102--2118},
	doi     = {10.1016/S0010-2180(01)00313-3}
}

@article{Leblanc2013PoF,
	author  = {Leblanc, Louis and Manoubi, Maha and Dennis, Kadeem and Liang, Zhe (Rita) and Radulescu, Matei I.},
	title   = {Dynamics of Unconfined Spherical Flames: Influence of Buoyancy},
	journal = {Physics of Fluids},
	year    = {2013},
	volume  = {25},
	number  = {9},
	pages   = {091106},
	doi     = {10.1063/1.4820018}
}

@article{Tse2000PCI,
	author  = {Tse, S. D. and He, L. and Law, C. K.},
	title   = {A Computational Study of the Transition from Localized Ignition to Flame Ball in Lean Hydrogen/Air Mixtures},
	journal = {Proceedings of the Combustion Institute},
	year    = {2000},
	volume  = {28},
	number  = {2},
	pages   = {1917--1924},
	doi     = {10.1016/S0082-0784(00)80596-2}
}

@article{WilliamsGrcar2009,
	author  = {Williams, Forman A. and Grcar, Joseph F.},
	title   = {A Hypothetical Burning-Velocity Formula for Very Lean Hydrogen--Air Mixtures},
	journal = {Proceedings of the Combustion Institute},
	year    = {2009},
	volume  = {32},
	number  = {1},
	pages   = {1351--1357},
	doi     = {10.1016/j.proci.2008.07.004}
}

@article{FernandezTarrazo2012IJHE,
	author  = {Fern{\'a}ndez-Tarrazo, Eduardo and S{\'a}nchez, Antonio L. and Li{\~n}{\'a}n, Amable and Williams, Forman A.},
	title   = {Flammability Conditions for Ultra-Lean Hydrogen Premixed Combustion Based on Flame-Ball Analyses},
	journal = {International Journal of Hydrogen Energy},
	year    = {2012},
	volume  = {37},
	number  = {2},
	pages   = {1813--1825},
	doi     = {10.1016/j.ijhydene.2011.10.037}
}

@article{makhviladze1998csat,
	title = {Numerical {{Modelling}} of {{Fireballs}} from {{Vertical Releases}} of {{Fuel Gases}}},
	author = {Makhviladze, G. M. and Roberts, J. P. and Yakush, S. E.},
	year = 1998,
	month = feb,
	journal = {Combustion Science and Technology},
	volume = {132},
	number = {1-6},
	pages = {199--223},
	issn = {0010-2202, 1563-521X},
	doi = {10.1080/00102209808952015},
	urldate = {2025-08-02},
	langid = {english}
}

@article{makhviladze2002potci,
	title = {Large-Scale Unconfined Fires and Explosions},
	author = {Makhviladze, G.M. and Yakush, S.E.},
	year = 2002,
	month = jan,
	journal = {Proceedings of the Combustion Institute},
	volume = {29},
	number = {1},
	pages = {195--210},
	issn = {15407489},
	doi = {10.1016/S1540-7489(02)80028-1},
	urldate = {2025-08-02},
	langid = {english}
}

@book{Kuo, 
	title={Fundamentals of turbulent and multiphase combustion},
	author={Kuo, K. K. and Acharya, R.},
	publisher={John Wiley \& Sons, Inc.},
	year={2012},
	edition={1},
	isbn={978-0-470-22622-3},
	totalpages={879},
}

@article{Baum1978,
	author = {Baum, H.R. and Rehm, R.G.},
	journal = {Journal of Research of the NBS},
	number = {3},
	pages = {297--308},
	title = {{The equations of motion for thermally driven, buoyant flows}},
	volume = {83},
	year = {1978}
}

@techreport{McGrattan,
	author = {McGrattan, Kevin and McDermott, Randall and Hostikka, Simo and Floyd, Jason and Vanella, Marcos},
	doi = {10.6028/NIST.SP.1018},
	institution={{U.S. Department of Commerce, National Institute of Standards and Technology}},
	publisher = {U.S. Department of Commerce},
	title = {{Fire Dynamics Simulator Technical Reference Guide Volume 1: Mathematical Model}},
	number={NIST Special Publication 1018-1},
	year = {2019},
	address={Gaithersburg, MD}
}

@article{yakovenko2024aa,
	title = {Ultra-Lean Hydrogen-Air Flames Initiated by a Hot Surface},
	author = {Yakovenko, I. and Melnikova, K. and Kiverin, A.},
	year = 2024,
	month = dec,
	journal = {Acta Astronautica},
	volume = {225},
	pages = {218--226},
	issn = {00945765},
	doi = {10.1016/j.actaastro.2024.09.013},
	urldate = {2025-01-12},
	langid = {english}
}

@article{tereza2023rjpcba,
	title = {Structure of a {{Lean Laminar Hydrogen}}–{{Air Flame}}},
	author = {Tereza, A. M. and Agafonov, G. L. and Anderzhanov, E. K. and Betev, A. S. and Medvedev, S. P. and Khomik, S. V. and Cherepanova, T. T.},
	date = {2023-08},
	journal = {Russian Journal of Physical Chemistry B},
	shortjournal = {Russ. J. Phys. Chem. B},
	volume = {17},
	number = {4},
	pages = {974--978},
	issn = {1990-7931, 1990-7923},
	doi = {10.1134/S1990793123040309},
	url = {https://link.springer.com/10.1134/S1990793123040309},
	urldate = {2026-01-16},
	langid = {english}
}

@article{tereza2025hfacp,
	title = {Influence of the Choice of Kinetic Mechanism on Predicted Structure of Lean Hydrogen--Air Flames},
	author = {Tereza, A. M.},
	year = 2025,
	journal = {Khimicheskaya Fizika / Advances in Chemical Physics},
	volume = {44},
	number = {4},
	pages = {79},
	issn = {3034-6126},
	doi = {10.7868/S3034612625040097},
	urldate = {2026-08-03},
	langid = {english}
}
	
	\PublishersNote{}
\end{document}